\documentclass[aps,twocolumn,floatfix,titlepage,balancelastpage,raggedbottom,amsfonts,amssymb,amsmath,showkeys]{revtex4}  
\usepackage{epsfig}
\usepackage{bm}
\usepackage{times}
\usepackage{mathptmx}
\usepackage{amsmath}
\usepackage{graphicx}
\usepackage{graphics}
\usepackage{amssymb}
\usepackage{bm}
\usepackage[english]{babel}
\begin{document}
\title{Surface growth effects on reactive capillary-driven flow: Lattice Boltzmann investigation}
\author{Danilo Sergi}
\affiliation{University of Applied Sciences (SUPSI), 
The iCIMSI Research Institute, 
Galleria 2, CH-6928 Manno, Switzerland}
\author{Loris Grossi}
\affiliation{University of Applied Sciences (SUPSI), 
The iCIMSI Research Institute, 
Galleria 2, CH-6928 Manno, Switzerland}
\author{Tiziano Leidi}
\affiliation{University of Applied Sciences (SUPSI), 
The iCIMSI Research Institute, 
Galleria 2, CH-6928 Manno, Switzerland}
\author{Alberto Ortona}
\affiliation{University of Applied Sciences (SUPSI), 
The iCIMSI Research Institute, 
Galleria 2, CH-6928 Manno, Switzerland}

\date{\today}

\begin{abstract}
The Washburn law has always played a critical role for ceramics. In the
microscale, surface forces take over volume forces and the phenomenon of 
spontaneous infiltration in narrow interstices becomes of particular relevance.
The Lattice Boltzmann method is applied in order to ascertain the role of 
surface reaction and subsequent deformation of a single capillary in 2D for the
linear Washburn behavior. The proposed investigation is motivated by
the problem of reactive infiltration of molten silicon into carbon preforms. 
This is a complex phenomenon arising from the interplay between fluid flow, the transition
to wetting, surface growth and heat transfer. Furthermore, it is characterized by slow 
infiltration velocities in narrow interstices resulting in small Reynolds numbers that are 
difficult to reproduce with a single capillary. In our simulations, several geometric 
characteristics for the capillaries are considered, as well as different infiltration and 
reaction conditions. The main result of our work is that the phenomenon of pore closure can be 
regarded as independent of the infiltration velocity, and in turn a number of other parameters. 
The instrumental conclusion drawn from our simulations is that short pores with wide openings and a 
round-shaped morphology near the throats represent the optimal configuration for the 
underlying structure of the porous preform in order to achieve faster infiltration.
The role of the approximations is discussed in detail and the robustness of our findings is
assessed.
\end{abstract}
\maketitle

\section*{1.~~~Introduction}

The infiltration of porous carbon (C) by molten silicon (Si) is a widespread industrial technique for the fabrication of composite materials with enhanced thermal and mechanical resistance at high temperatures (Hillig et al., 1975). Central to this process are of course the wetting properties of the substrate against molten Si. Importantly, it appears that capillary forces are strongly affected by the reactivity at high temperatures between Si and C to form silicon carbide (SiC). A large body of research has been devoted to the subject, definitely defying conventional modeling (Dezellus and Eustathopoulos, 2010; Dezellus et al., 2003; Mortensen et al., 1997) and experimental (Eustathopoulos et al., 1999) approaches. The interested reader can find more details about the experimental work in Liu et al.~(2010) and references therein. Our simulations are based on the Lattice Boltzmann (LB) method (Benzi et al., 1992; Succi, 2009; Sukop and Thorne, 2010). For advanced capillary problems, the accuracy of this numerical scheme for hydrodynamics has been proven to be comparable with the well-established method of Computational Fluid Dynamics (CFD) relying on finite elements (Chibbaro et al., 2009b). Here we deal with the ability of the LB method  to address the dynamic behavior of capillary infiltration coupled to surface growth. Of special interest is the retardation induced to the Washburn law by the dynamic reduction in pore radius. More generally, the present work can help to understand experimental results carried out with Si alloys (Bougiouri et al., 2006; Calderon et al., 2010a; Sangsuwan et al., 1999; Voytovych et al., 2008). Precisely, the focus is on the linear Washburn law for a single capillary in 2D (Chibbaro, 2008; Chibbaro et al., 2009a). This behavior is typical for the infiltration of pure Si in carbon preforms (Israel et al., 2010). Furthermore, the Washburn analysis for bulk structures is based on the results for a single capillary (Einset, 1996 and 1998). In the LB model, the surface reaction is treated as a precipitation process from supersaturation (Bougiouri et al., 2006; Calderon et al., 2010a and 2010b; Kang et al., 2007, 2003, 2002b, 2004; Lu et al., 2009). It is also worth mentioning the article by Miller and Succi (2002) as pioneering work using the LB method for this subject. Our simulations indicate that the thickening of the surface behind the contact line actually limits the infiltration process. This phenomenon is especially marked in the vicinity of the onset of pore closure. For this reason, we argue that the effect of retardation is stronger when surface growth inside narrow interstices obstructs the flow (here the wetting transition at the contact line is neglected (Calderon et al., 2010b; Dezellus et al., 2005; Voytovych et al., 2008)). Importantly, it arises that higher infiltration velocities, achieved for example by tuning some parameters defining the capillary geometry, have little to no influence on the process of pore closure by surface reaction. It follows that the distribution of pore size is a critical parameter for reactive capillary infiltration, besides other aspects to be discussed in the sequel.The engineering problem under consideration is an involved process arising from the mutual dependence between various phenomena. Unavoidably, our description suffers from several simplifications. We also verify that our findings remain robust under the existing approximations.


\section*{2.~~~Computational models}\label{sec:models}

The distinctive feature of the LB method resides in the discretization of the
velocity space: only a finite number of elementary velocity modes is available.
Locally, the movement of the fluid particles along a particular direction is accounted
for statistically by means of distribution functions, in terms of which the
intervening physical quantities are expressed. Hydrodynamic behavior is
recovered in the limit of small Mach numbers, corresponding to the incompressible limit (Chen and Doolen, 1998). 
It has been shown that the problem of capillary flow is better described by a multicomponent
system (Kang et al., 2002a; Shan and Chen, 1993; Shan and Doolen, 1995), since allowing to reduce the evaporation-condensation
effect (Chibbaro, 2008; Chibbaro et al., 2009a; Diotallevi et al., 2009a and 2009b; Pooley et al., 2009). On a d2q9 lattice, the evolution 
of every single fluid component is obtained by iterating the BGK equation (Bhatnagar et al., 1954)
\begin{multline}
f^{\sigma}_{i}(\bm{r}+\bm{e}_{i}\Delta t,t+\Delta t)=\\
f^{\sigma}_{i}(\bm{r},t)-\frac{1}{\tau_{\sigma}}\big[f^{\sigma}_{i}(\bm{r},t)-
f_{i}^{\sigma,\mathrm{eq}}(\rho_{\sigma},\bm{u}_{\sigma}^{\mathrm{eq}})\big]\quad i=0-8\ .
\label{eq:bgk}
\end{multline}
The lattice spacing is denoted by $\Delta x$ and the time increment by $\Delta t$; hereafter, without
loosing generality $\Delta x=\Delta t=1$ (in model units). The $f^{\sigma}_{i}$'s are the distribution 
functions for the velocity modes $\bm{e}_{i}$. The superscript $\sigma=1,2$ designates the substance.  
The discrete velocity modes are defined as follows
\[
\bm{e}_{i}=
\left\{\begin{array}{ll}
(0,0) & i=0\\
(\cos[(i-1)\pi/2],\sin[(i-1)\pi/2]) & i=1-4\\
\sqrt{2}(\cos[(2i-9)\pi/4],\sin[(2i-9)\pi/4]) & i=5-8\ .\\
\end{array}
\right.\]
The equilibrium distribution functions read
\begin{multline*}
f^{\sigma,\mathrm{eq}}_{i}(\rho_{\sigma},\bm{u}_{\sigma}^{\mathrm{eq}})=w_{i}\rho_{\sigma}(\bm{r},t)\Big[1
+3\bm{e}_{i}\cdot\bm{u}_{\sigma}^{\mathrm{eq}}
\\+\frac{9}{2}(\bm{e}_{i}\cdot\bm{u}_{\sigma}^{\mathrm{eq}})^{2}
-\frac{3}{2}\bm{u}_{\sigma}^{\mathrm{eq}}\cdot\bm{u}_{\sigma}^{\mathrm{eq}}\Big]\quad i=0-8\ .
\end{multline*}
The coefficients $w_{i}$ are defined as $w_{0}=4/9$, $w_{i}=1/9$ for $i=1-4$ and $w_{i}=1/36$
for $i=5-8$ (He and Luo, 1997; Wolf-Gladrow, 2005). The local density, $\rho_{\sigma}$, and the equilibrium velocity, 
$\bm{u}^{\mathrm{eq}}_{\sigma}$, have the explicit form
\begin{eqnarray*}
&&\rho_{\sigma}(\bm{r},t)=\sum_{i=0}^{8}f^{\sigma}_{i}(\bm{r},t)\\ 
&&\bm{u}_{\sigma}^{\mathrm{eq}}(\bm{r},t)=\bm{u}'(\bm{r},t)
+\frac{\tau_{\sigma}\bm{F}_{\sigma}(\bm{r},t)}{\rho_{\sigma}(\bm{r},t)}\ .
\end{eqnarray*}
$\bm{F}_{\sigma}$ is the resultant force experienced by the fluid component $\sigma$.
The velocity $\bm{u}'$ is given by
\begin{equation*}
\bm{u}'=\Big(\sum_{\sigma}\frac{\rho_{\sigma}(\bm{r},t)\bm{u}_{\sigma}(\bm{r},t)}{\tau_{\sigma}}\Big)\bigg/
\sum_{\sigma}\frac{\rho_{\sigma}(\bm{r},t)}{\tau_{\sigma}}\ ,
\end{equation*}
where we have introduced the $\sigma$-th component fluid velocity
\begin{equation*}
\bm{u}_{\sigma}(\bm{r},t)=\frac{1}{\rho_{\sigma}(\bm{r},t)}\sum_{i}f_{i}^{\sigma}(\bm{r},t)\bm{e}_{i}\ .
\end{equation*}
The velocity $\bm{u}'$ satisfies the requirement of momentum conservation in the absence
of external forces (Kang et al., 2002a; Shan and Chen, 1993; Shan and Doolen, 1995).
In Eq.~\ref{eq:bgk}, $\tau_{\sigma}$ is the relaxation time, determining the kinematic viscosity 
$\nu_{\sigma}=(2\tau_{\sigma}-1)/6$. The dynamic viscosity is defined as (Chibbaro, 2008)
\begin{equation*}
\mu=\rho\nu=\sum_{\sigma}\mu_{\sigma}=\sum_{\sigma}\rho_{\sigma}\nu_{\sigma}\ .
\end{equation*}
The total density of the whole fluid is of course $\rho=\sum_{\sigma}\rho_{\sigma}$.
Collisions between the liquid phase and the solid boundaries are treated according to the
common bounce-back rule, implementing the no-slip condition for rough surfaces. Cohesive 
forces in the liquid phase are evaluated by means of the formula (Martys and Chen, 1996)
\begin{equation*}
\bm{F}_{\mathrm{c},\sigma}(\bm{r},t)=-G_{\mathrm{c}}\rho_{\sigma}(\bm{r},t)\sum_{i=1}^{8}w_{i}
\rho_{\bar{\sigma}}(\bm{r}+\bm{e}_{i}\Delta t,t)\bm{e}_{i}\ ;
\end{equation*}
$\bar{\sigma}$ being the other fluid component with respect to $\sigma$.
The parameter $G_{\mathrm{c}}$ determines the interaction strength. Adhesive forces between the liquid
and solid phases are computed using the formula (Martys and Chen, 1996)
\begin{equation*}
\bm{F}_{\mathrm{ads},\sigma}(\bm{r},t)=-G_{\mathrm{ads},\sigma}\rho_{\sigma}(\bm{r},t)\sum_{i=1}^{8}w_{i}
s(\bm{r}+\bm{e}_{i}\Delta t,t)\bm{e}_{i}\ .
\end{equation*}
The interaction strength is adjusted via the parameter $G_{\mathrm{ads},\sigma}$. The function $s$ takes
on the value $1$ if the velocity mode $\bm{e}_{i}$ points to a solid node, otherwise it vanishes.

The LB scheme can encompass convection-diffusion systems readily under the assumption of low
concentrations. This condition implies that fluid flow is not affected by molecular
diffusion. Solute transport is described by introducing the distribution functions $g_{i}$.
For solute transport we use the d2q4 lattice (Kang et al., 2007), generally applied also for thermal fields 
(Mohamad et al., 2009; Mohamad and Kuzmin, 2010). The reduced number of degrees of freedom does not restrain the predictive 
capacity of the model (Wolf-Gadrow, 1995).  As for fluid flow, the evolution follows the 
BGK equation (Kang et al., 2007, 2003, 2002b, 2004; Lu et al., 2009)
\begin{multline*}
g_{i}(\bm{r}+\bm{e}_{i}\Delta t,t+\Delta t)=g_{i}(\bm{r},t)
-\frac{1}{\tau_{\mathrm{s}}}\big[g_{i}(\bm{r},t)-g_{i}^{\mathrm{eq}}(C,\bm{u})\big]\\
i=1-4\ .
\end{multline*}
The relaxation time for solute transport is denoted $\tau_{\mathrm{s}}$, defining
the diffusion coefficient $D=(2\tau_{\mathrm{s}}-1)/4$ (Kang et al., 2007). For the d2q4 lattice,
the equilibrium distribution functions take the form (Kang et al., 2007)
\begin{equation}
g_{i}^{\mathrm{eq}}(C,\bm{u})=\frac{1}{4}C\big[1+2\bm{e}_{i}\cdot\bm{u}\big]
\quad i=1-4\ .
\label{eq:ueq}
\end{equation}
The solute concentration is given by $C=\sum_{i}g_{i}$. The local solvent velocity $\bm{u}$ is assumed
to be the overall fluid velocity (Kang et al., 2002a; Shan and Chen, 1993; Shan and Doolen, 1995)
\begin{multline*}
\bm{u}(\bm{r},t)=\Big(\sum_{\sigma}\Big[\rho_{\sigma}(\bm{r},t)\bm{u}_{\sigma}(\bm{r},t)\\
+\frac{1}{2}\bm{F}_{\sigma}(\bm{r},t)\Big]\Big)\bigg/\sum_{\sigma}\rho_{\sigma}(\bm{r},t)\ .
\end{multline*}

It may be useful to have interfaces also for solute transport in order to prevent diffusion
from the liquid to the vapor phase. In analogy with multiphase fluid flow (Shan and Chen, 1993 and 1994), 
we add to the velocity $\bm{u}$ in Eq.~\ref{eq:ueq} a term of the form $\tau_{\mathrm{s}}\bm{F}_{\mathrm{s}}/C$.
The function $\bm{F}_{\mathrm{s}}$ is defined as
\begin{equation}
\bm{F}_{\mathrm{s}}(\bm{r},t)=-G_{\mathrm{s}}\varphi(\bm{r},t)\sum_{i=1}^{4}\varphi(\bm{r}+\bm{e}_{i}\Delta t,t)
\bm{e}_{i}\ .
\label{eq:fs}
\end{equation}
The parameter $G_{\mathrm{s}}$ allows to tune the intensity. The function $\varphi(\bm{r},t)$
is given by
\begin{equation*}
\varphi(\bm{r},t)=\varphi_{0}e^{-C_{0}/C(\bm{r},t)}\ .
\end{equation*}
This mechanism introduces an interface acting as a barrier to diffusion moving with the
overall fluid. Far from interfaces, $\bm{F}_{\mathrm{s}}$ vanishes and diffusive transport is
restored. The contribution of $\bm{F}_{\mathrm{s}}$ is taken into account also for the nodes
belonging to the surface of the solid phase.

At the reactive boundary, the solute concentration $C$ satisfies the macroscopic
equation (Kang et al., 2007, 2003, 2002b, 2004; Lu et al., 2009)
\begin{equation*}
D\frac{\partial C}{\partial n}=k_{\mathrm{r}}\big(C-C_{\mathrm{s}}\big)\ ,
\end{equation*}
where $k_{\mathrm{r}}$ is the reaction-rate constant and $C_{\mathrm{s}}$ the
saturated concentration. From the constraint of the above equation, it is 
possible to calculate the solute concentration at the solid-liquid interface (Kang et al., 2007)
\begin{equation*}
C=\frac{2g_{\mathrm{out}}+k_{\mathrm{r}}C_{\mathrm{s}}}{k_{\mathrm{r}}+1/2}\ ,
\end{equation*}
where $g_{\mathrm{out}}$ is the distribution function of the modes leaving the fluid phase.
At the solid boundaries, the distribution functions $g_{\mathrm{in}}$ for the
modes $\bm{e}_{i}$ pointing in the fluid phase remain undetermined. They are updated
according to the rule $g_{\mathrm{in}}=-g_{\mathrm{out}}+C/2$; $g_{\mathrm{in}}$ and $g_{\mathrm{out}}$
are associated with opposite directions. As time goes on, solute mass deposits 
on the reacting solid surface. The mass is updated iterating the equation 
(Kang et al., 2004; Lu et al., 2009)
\begin{equation*}
b(\bm{r},t+\Delta t)=b(\bm{r},t)+k_{\mathrm{r}}(C-C_{\mathrm{s}})\ .
\end{equation*}
We indicate by $b_{0}$ the initial mass present on solid boundaries. The solid surface
grows whenever $b(\bm{r},t)$ exceeds the threshold value $b_{\mathrm{max}}$. 
The new solid node is chosen  with uniform probability among the surrounding nodes belonging
to the liquid phase. Only direct neighbors are taken into account. The cumulated mass of the
solid node that triggered surface growth is set to zero.

Lattice Boltzmann simulations process and output data in model units. The different quantities 
will be expressed in terms of the basic model units for mass, length and time, denoted by  mu,
lu and ts, respectively. Direct comparison with physical systems is possible after suitable 
transformations (Lu et al., 2009). This procedure is not strictly necessary since, in order to describe 
properly the real systems, it is sufficient to have the same Reynolds number $Re$  for fluid flow 
(Landau and Lifshitz, 2008), and the same Damkohler number $Da$ and supersaturation $\psi$ for solute transport
(Kang et al., 2003 and 2004; Lu et al., 2009). These dimensionless 
parameters are the same in LB and physical units, expressed for example in the common SI system
or the Gauss system. For the reactive infiltration of molten Si in 
a capillary of $10$ $\mu$m in width, the Reynolds number can be estimated to be 
$Re\approx 10^{-6}$ (Einset, 1996; Voytovych et al., 2008). This Reynolds number asks for extreme simulation 
conditions that can not be achieved. Indeed, a channel of width $50$ lu using $\tau_{\sigma}=1$ ts
would require a velocity of infiltration $u\approx 10^{-8}$ lu/ts. Thus, in the following,
it should be borne in mind that our simulations significantly overestimate the typical Reynolds
numbers for the problem at hand. In other words, this means that the process of infiltration
is too fast. This numerical deficiency is partly reduced by the fact that the precipitation process 
from supersaturation should not have a strong dependence from the velocity of the invading front 
(Kang et al., 2003 and 2004; Lu et al., 2009).


\section*{3.~~~Results and Discussion}

\subsection*{3.1~~~Interfaces and surface tension}\label{sec:gamma}

\begin{figure}[t]
\includegraphics[width=8.5cm]{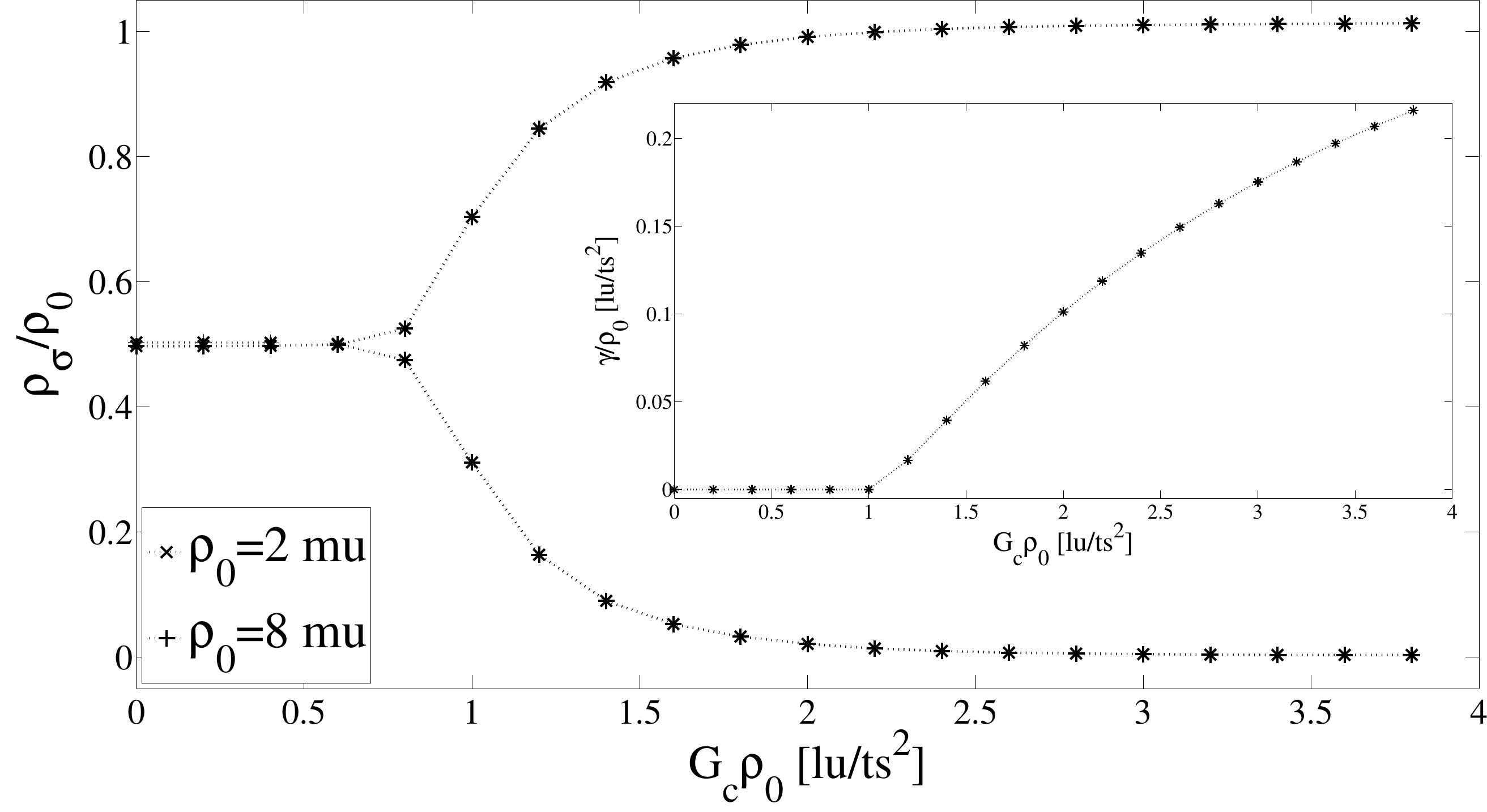}
\caption{\label{fig:density} Scaled density $\rho_{\sigma}/\rho_{0}$ as a function of
$G_{\mathrm{c}}\rho_{0}$. The upper branch corresponds to $\rho_{1}/\rho_{0}$ inside the
bubble (main density), while the lower branch is $\rho_{2}/\rho_{0}$ (dissolved density).
Inset: Dependence of the scaled surface tension $\gamma/\rho_{0}$ on $G_{\mathrm{c}}\rho_{0}$.
The results of the simulations are averaged over the last $10$ frames.}
\end{figure}

Cohesive forces among molecules of the same species have the effect to give rise to
interfaces separating the fluid components. Given two immiscible fluids, the work
necessary in order to deform the interface of area $A$ is $\delta W=\gamma\mathrm{d}A$,
where $\gamma$ is the surface tension. By taking into account the work of pressure
forces, the condition of mechanical equilibrium leads to Laplace law (de Gennes et al., 2004;
Kang et al, 2002a)
\begin{equation}
\Delta P=\frac{\gamma}{r}\quad\text{in 2D.}
\label{eq:laplace}
\end{equation}
In the above formula, $r$ is the radius of the spherical droplet. The pressure 
drop across the interface has the explicit form 
$\Delta P=P_{\mathrm{in}}-P_{\mathrm{out}}$, with self-explanatory notation. In order to understand 
the interplay between interfacial tension, wetting and capillary flow, we first consider 
the formation of stable droplets for systems containing two fluid components. The size 
of the simulation domain is $N_{x}\times N_{x}$ 
with $N_{x}=200$ lu; periodic boundary conditions are applied. In the center is placed 
a droplet of fluid $1$ immersed in fluid $2$. Inside the bubble, $\rho_{1}$ is the main 
density and $\rho_{2}$ is the dissolved density. For the initial condition, we assume that
$\rho_{2}/\rho_{1}=2.5\%$ and outside the bubble their values are inverted. The
initial total density is denoted by $\rho_{0}=\rho_{1}+\rho_{2}$. By choosing the 
initial radius of the droplet to be $r=N_{x}/\sqrt{2\pi}$, it follows that both fluids 
have equal total masses, as assumed by Huang et al.~(2007). The densities of both
fluids determine the pressure $P$. Let $\rho_{1}$ and $\rho_{2}$ be the densities
of fluids $1$ and $2$ at point $\bm{r}=(x,y)$, then the pressure is given by 
(Huang et al., 2007; Shan and Chen, 1993; Shan and Doolen, 1995)
\begin{equation}
P(\bm{r})=\frac{\rho_{1}(\bm{r})+\rho_{2}(\bm{r})}{3}
+\frac{G_{\mathrm{c}}}{3}\rho_{1}(\bm{r})\rho_{2}(\bm{r})\ .
\label{eq:pressure}
\end{equation}
In Eq.~\ref{eq:laplace}, for $P_{\mathrm{in}}$ we use the value obtained by averaging
over a square of side length $2L_{0}+1$ centered in the simulation domain; $L_{0}=20$ lu. 
Instead, for $P_{\mathrm{out}}$, the average is taken over the square $[0,2L_{0}+1]\times[0,2L_{0}+1]$.
Every dynamics amounts to $100'000$ timesteps. For the analysis, $50$
uniformly-spaced frames are collected. In order to derive the radius of the droplets, it 
is necessary to determine the location of the interface separating the fluid components.
To do this, we start from the point $(0,N_{y}/2)$ and consider 
at every step points incremented by one unit in $x$, i.e.~$x\rightarrow x+1$, applying every 
time the following rules. If $\rho_{1}(x+1)/\rho_{1}(x)>1.05$ holds, it follows that the 
density for the first fluid increases at least of $5\%$ in passing from $x$ to $x+1$. We 
consider these lattice sites as belonging to the interface. For this set of points, we 
determine $x_{\mathrm{max}}$ and $x_{\mathrm{min}}$ and it is assumed that the interface has 
coordinates $x=(x_{\mathrm{max}}+x_{\mathrm{min}})/2$ and $y=N_{y}/2$. Figure \ref{fig:density}
illustrates the results according to the analysis proposed by Huang et al.~(2007). It turns
out that the surface tension $\gamma$ is appreciable only for $G_{\mathrm{c}}\rho_{0}>1$, as
thoroughly discussed by Huang et al.~(2007). We consider outliers the two points for 
which $G_{\mathrm{c}}\rho_{0}\leq 1$ and exhibiting a weak phase separation (broad interface), since
inspection of the density values during the dynamics indicates that the systems are still
evolving towards the equilibrium state after $100'000$ timesteps. For $\rho_{0}=2$ mu/lu$^{2}$
and $G_{\mathrm{c}}\rho_{0}=1$ lu/ts$^{2}$, the average over the last $20'000$ timesteps
for a simulation of $500'000$ timesteps yields $\rho_{1}=1.17744$ mu/lu$^{2}$. For the shorter
dynamics we obtained $\rho_{1}=1.40671$ mu/lu$^{2}$. Furthermore, there appears that, even
for the longer simulation, the density values still exhibit systematic variations of
at least $1\%$ over the last $20'000$ timesteps. It is thus clear that $\rho_{1}/\rho_{0}$ tends
towards $0.5$ (Huang et al., 2007). For $1'000'000$ iterations this limit is further approached
since $\rho_{1}=1.12514$ mu/lu$^{2}$. On the other hand, for $\rho_{0}=2$ mu/lu$^{2}$ and 
$G_{\mathrm{c}}\rho_{0}=1.4$ lu/ts$^{2}$, we obtain $\rho_{1}=1.83668$ mu/lu$^{2}$ in the case
of a long simulation. The results remain practically unchanged since for the shorter
simulation we have $\rho_{1}=1.83733$ mu/lu$^{2}$ and the difference between the two corresponding
values for the surface tension is on the order of $10^{-5}$ mu/lu$^{2}$. For systems with higher surface
tension the convergence towards the equilibrium state is faster. 
In the following, we will concentrate ourselves on three particular values of surface 
tension, leading in general to consistent and quite accurate results.


\subsection*{3.2~~~Contact angle}\label{sec:theta}

\begin{figure}[t]
\includegraphics[width=8.5cm]{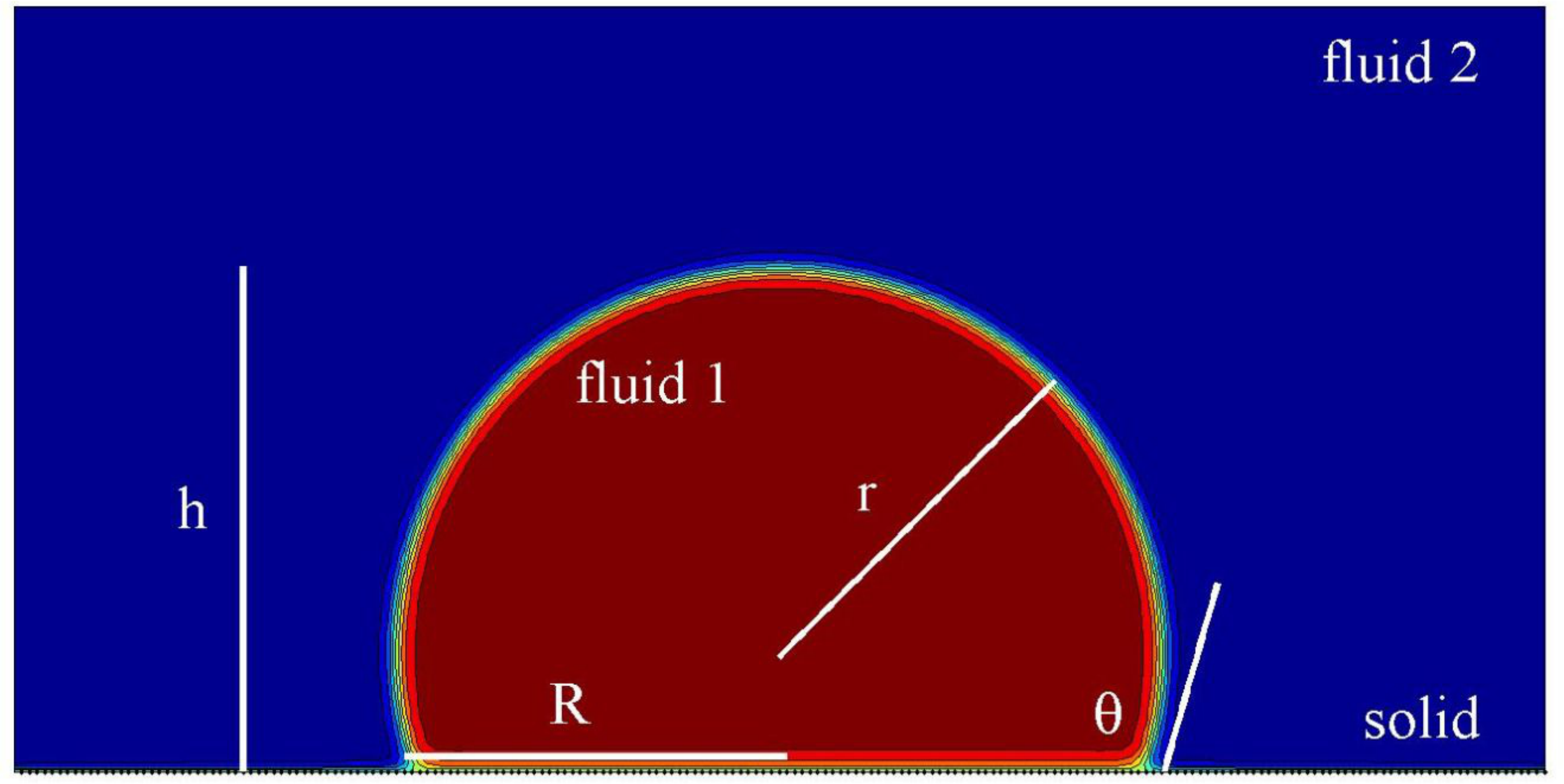}
\caption{\label{fig:droplet} Sessile droplet system for the determination of
the contact angle $\theta$. The solid line is the tangent at the triple line. 
Color code based on density. $h$ is the height, $r$ the radius and $R$ the
base radius.}
\end{figure}

The wetting behavior of a surface against a liquid is usually characterized
via the equilibrium contact angle. Given the profile of the droplet, it is
defined as the angle formed by the tangent at the contact line, that is, the
edge where the phases solid (S), liquid (L) and vapor (V) meet (see Fig.~\ref{fig:droplet}).
For macroscopic, spherical droplets, Young equation (Young, 1805) relates the 
contact angle $\theta$ to the interfacial tensions $\gamma_{ij}$ ($i,j=$S,L,V):
\begin{equation*}
\cos\theta=\frac{\gamma_{\mathrm{SV}}-\gamma_{\mathrm{SL}}}{\gamma}\ .
\end{equation*}
For the solid-liquid surface tension we continue to use the simpler
notation $\gamma$. If $\theta<90^{\circ}$, the liquid wets the substrate
(hydrophilic regime); otherwise, if $\theta>90^{\circ}$, the substrate is
hydrophobic. 

In our simulations, the contact angle is determined using
the sessile droplet method. The simulation domain is  $N_{x}=400$ lu long and
$N_{y}=200$ lu wide. The substrate coincides with the bottom of the 
simulation domain; the other boundaries are periodic. In the initial state, a square 
of side length $80$ lu of the first fluid
is centered in the simulation domain, placed in contact with the solid
phase. The systems consist of a binary mixture where the droplet is
represented by the wetting fluid immersed in the non-wetting component.
As in Sec.~3.1, for the initial condition, $\rho_{1}/\rho_{2}=2.5\%$
with $\rho_{0}=\rho_{1}+\rho_{2}=2$ mu/lu$^{2}$. The simulations are
performed with three different values for $G_{\mathrm{c}}$. The equilibrium contact 
angle is varied by considering different values for the interaction parameters 
$G_{\mathrm{ads,\sigma}}$.  The evolution of the systems consists of $100'000$
timesteps.

\begin{figure}[t]
\includegraphics[width=8.5cm]{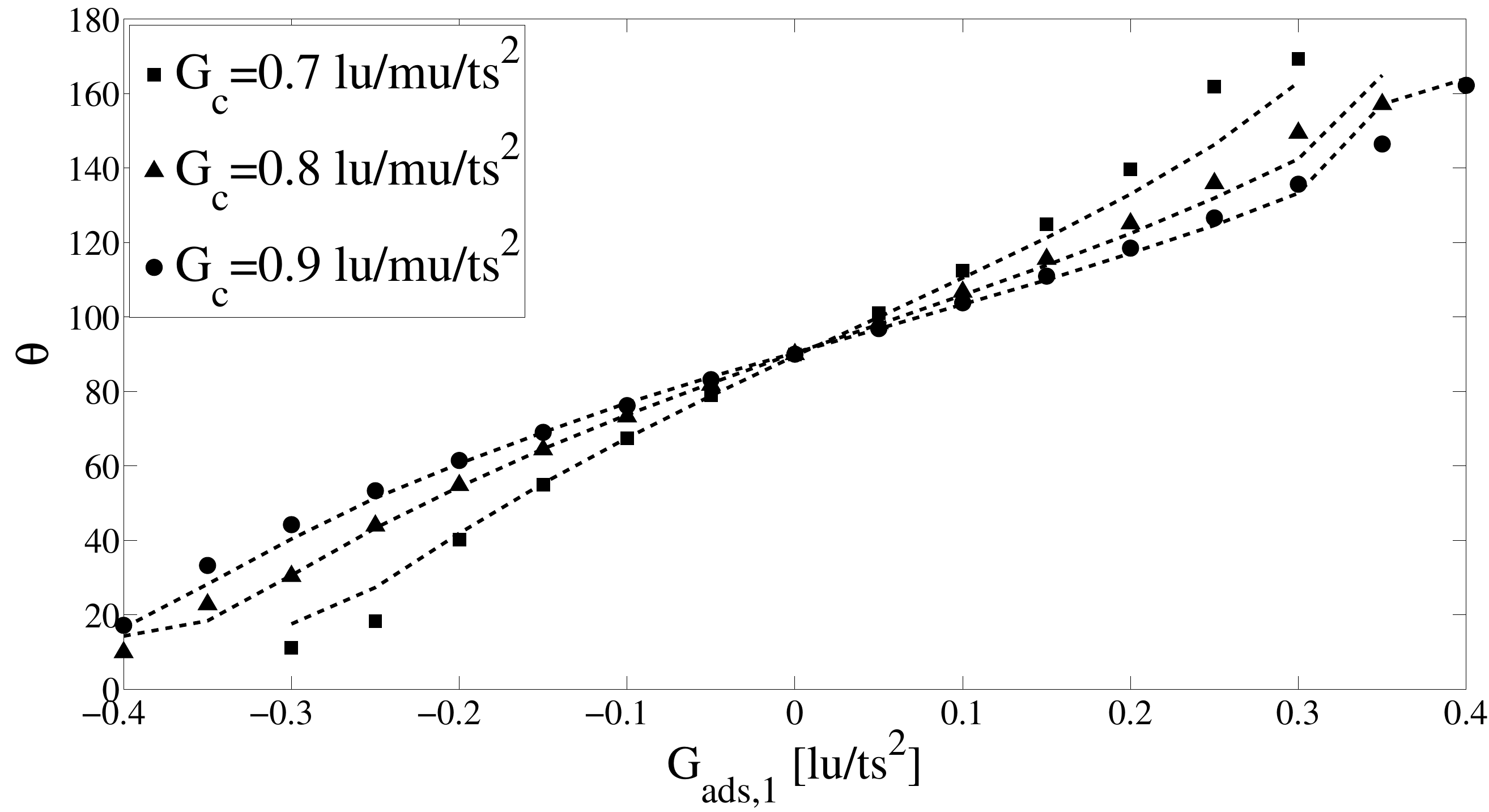}
\caption{\label{fig:theta} Dependence of the contact angle $\theta$ on the solid-liquid
interaction parameter $G_{\mathrm{ads},1}=-G_{\mathrm{ads},2}$ for the simulation of the sessile
droplet system (see Fig.~\ref{fig:droplet}). Three different surface tensions are considered by 
varying the parameter $G_{\mathrm{c}}$ (see Fig.~\ref{fig:density}). Points represent the results 
directly obtained from the profiles. The results predicted by Eq.~\ref{eq:huang} correspond to 
the dashed lines. The results of the simulations are averaged over the last $10$ frames. The 
cases for which it is difficult to extract the contact angle are omitted.}
\end{figure}

In the analysis, the contact angle is extracted using the method employed by 
Schmieschek and Harting (2011). In the micron scale, the droplet profile can be 
safely assumed to be spherical (Sergi et al., 2012). Here, our systems 
refer to millimeter-sized droplets (Bougiouri et al., 2006; Voytovych et al., 2008). So, 
let $h$ be the height of the spherical cap and $R$ the base radius (contact line curvature), 
then the radius and the contact angle can be written respectively as
\begin{equation*}
r=\frac{h^{2}+R^{2}}{2h}\quad\text{and}\quad
\theta=\frac{\pi}{2}\mp\arccos\Big(\frac{R}{r}\Big)\ ,
\end{equation*}
for the hydrophilic ($R/h>1$) and hydrophobic ($R/h<1$) regimes, respectively. 
For clarity, these quantities are shown in Fig.~\ref{fig:droplet}.
For the determination of the height and base radius of the droplets the interface
between the wetting and non-wetting fluid is derived as explained in Sec.~3.1. 
An alternative way for calculating the contact angle is to use  the formula (Huang et al., 2007)
\begin{equation}
\cos\theta=\frac{G_{\mathrm{ads},2}-G_{\mathrm{ads},1}}{
G_{\mathrm{c}}\frac{\rho_{1}-\rho_{2}}{2}}\ .
\label{eq:huang}
\end{equation}
$\rho_{1}$ and $\rho_{2}$ are the main and dissolved densities averaged over the region
$[0,2L_{0}+1]\times[0,2L_{0}+1]$ with $L_{0}=5$ lu centered in the point $(N_{x}/2,h/2)$.
Figure \ref{fig:theta} plots the contact angle versus the interaction parameter controlling
adhesive forces between the solid and liquid phases. As the surface tension increases, it
appears that the agreement between the two approaches is more stringent. Our findings are in good 
accordance with those reported by Huang et al.~(2007).


\begin{figure*}[t]
\includegraphics[width=12cm]{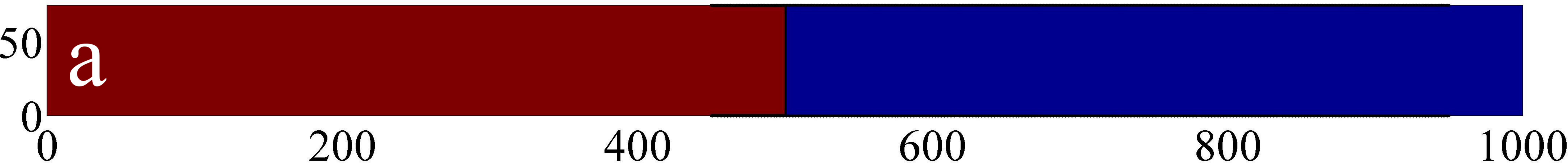}\\
\includegraphics[width=12cm]{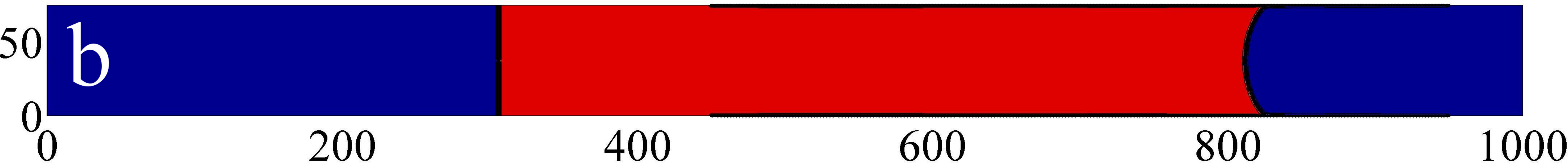}\\
\includegraphics[width=12cm]{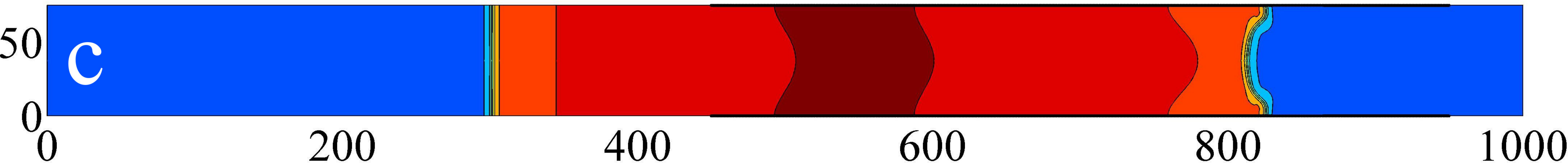}
\caption{\label{fig:capillary}
Set-up for the simulations of capillary-driven flow. Color code based on density: the first fluid is 
represented in red (high) and the second one in blue (low). Black points represent the solid boundary. 
The simulation domain has periodic boundary conditions. (a) Initial configuration. 
The first fluid fills the capillary along a distance of $50$ lu. (b) State of the system after the 
formation of the meniscus and partial penetration. The flat interface guarantees that the first fluid can 
be regarded as forming an infinitely large phase, mimicking in this way a reservoir. (c) In this case 
the color code is based on solute concentration. The region filled with the first fluid is at a higher concentration
and the dividing interface from the low-concentration region approximately coincides with the interface
for fluid flow.}
\end{figure*}
\begin{figure*}[t]
\includegraphics[width=12cm]{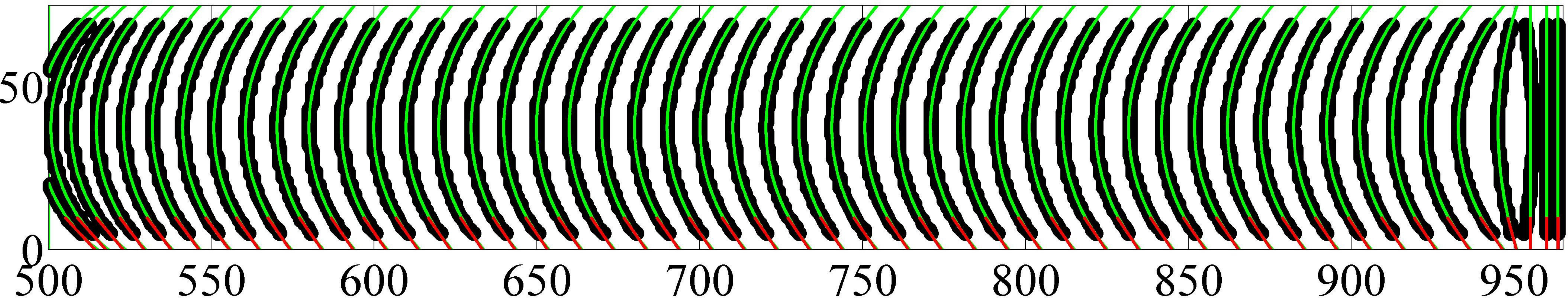}
\caption{\label{fig:front} 
Infiltration front at different times. Black points represent the profile as obtained from
simulations. The green line is a fit to the data using a circle as model function. The red straight
line is the tangent at the point where the contact angle $\theta$ for the wetting fluid is determined.
In general, the results determined in this way are in reasonable accordance with those based on  
capillary pressure (cf.~Chibbaro et al.~(2009b)).}
\end{figure*}
\begin{figure}[t]
\includegraphics[width=8.5cm]{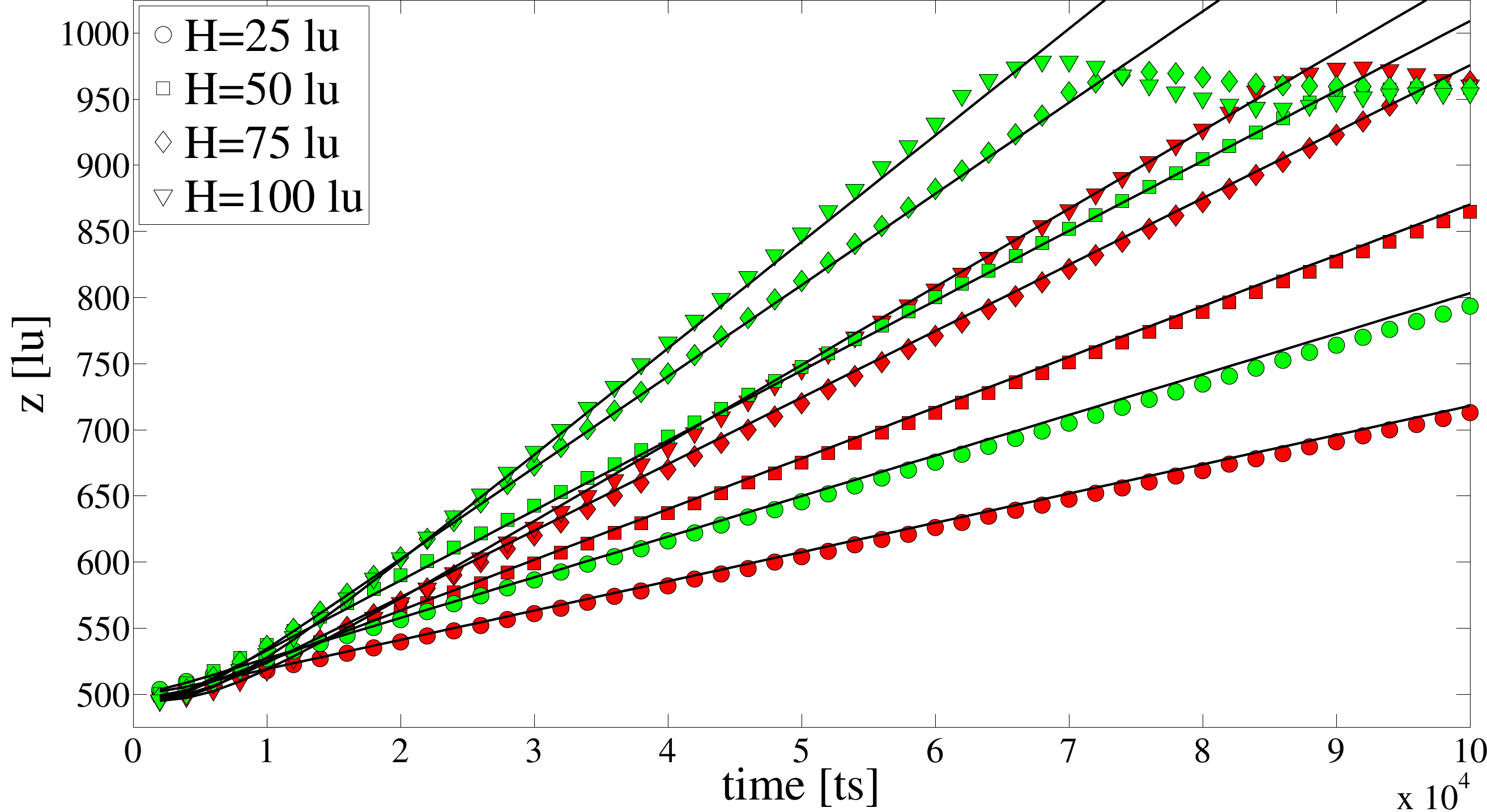}
\caption{\label{fig:washburn} 
Liquid displacement inside the capillary versus time. Points represent the 
simulation results while the solid lines correspond to the theoretical prediction of
Eq.~\ref{eq:washburn}. The data shown in red are obtained with the surface tension 
$\gamma=0.12338$ lu$\cdot$mu/ts$^{2}$ ($G_{\mathrm{c}}=0.8$ lu/mu/ts$^{2}$). For the results displayed in green 
the surface tension is $\gamma=0.16403$ lu$\cdot$mu/ts$^{2}$ ($G_{\mathrm{c}}=0.9$ lu/mu/ts$^{2}$). 
The length of the channels is $L=500$ lu. The results of the simulations are averaged over $10$ frames.}
\end{figure}
\begin{figure}[t]
\includegraphics[width=8.5cm]{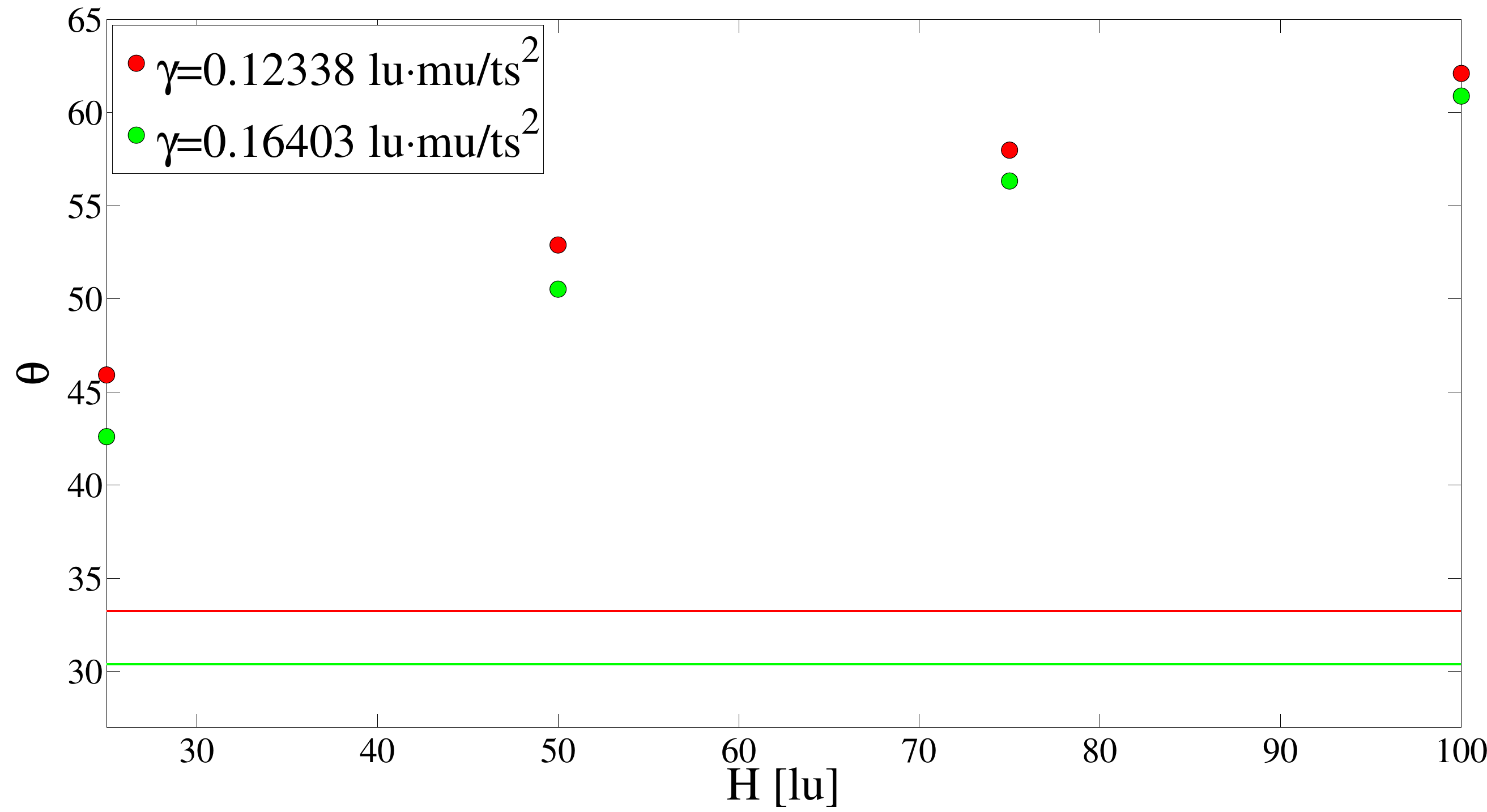}
\caption{\label{fig:thetavsH} 
Dynamic contact angle, average value, as a function of the capillary width $H$ (see Eq.~\ref{eq:washburn}).
The solid lines represent the equilibrium contact angle as obtained using the sessile droplet system 
(see Sec.~3.2). The results are for channels of length $L=500$ lu.}
\end{figure}
\begin{figure*}[t]
\includegraphics[width=12cm]{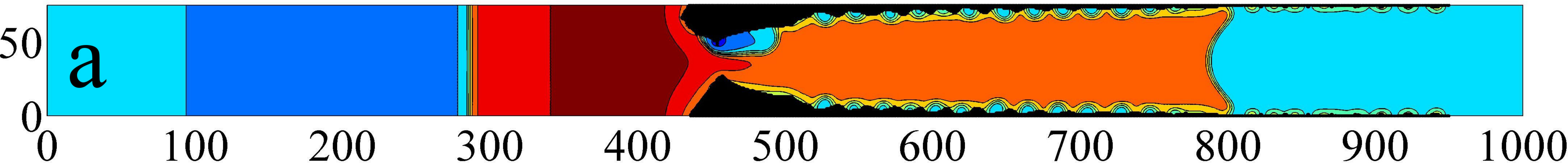}\\
\includegraphics[width=12cm]{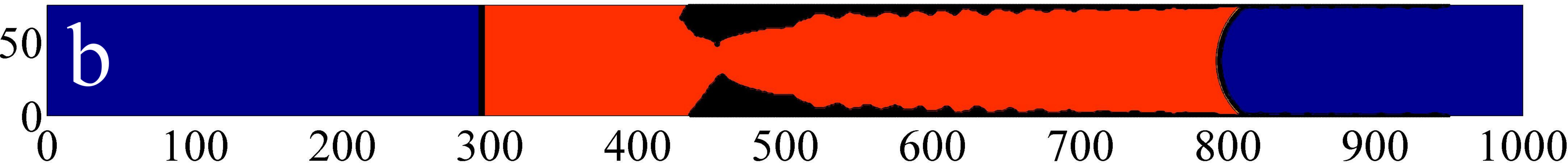}
\caption{\label{fig:growth}
Typical interstice morphology as resulting from surface growth. The solid phase is represented in black.
In red tonality the main component for mass transport and fluid flow. (a) Representation for solute
concentration. (b) Representation for fluid density.}
\end{figure*}
\begin{figure*}[t]
\includegraphics[width=8.5cm]{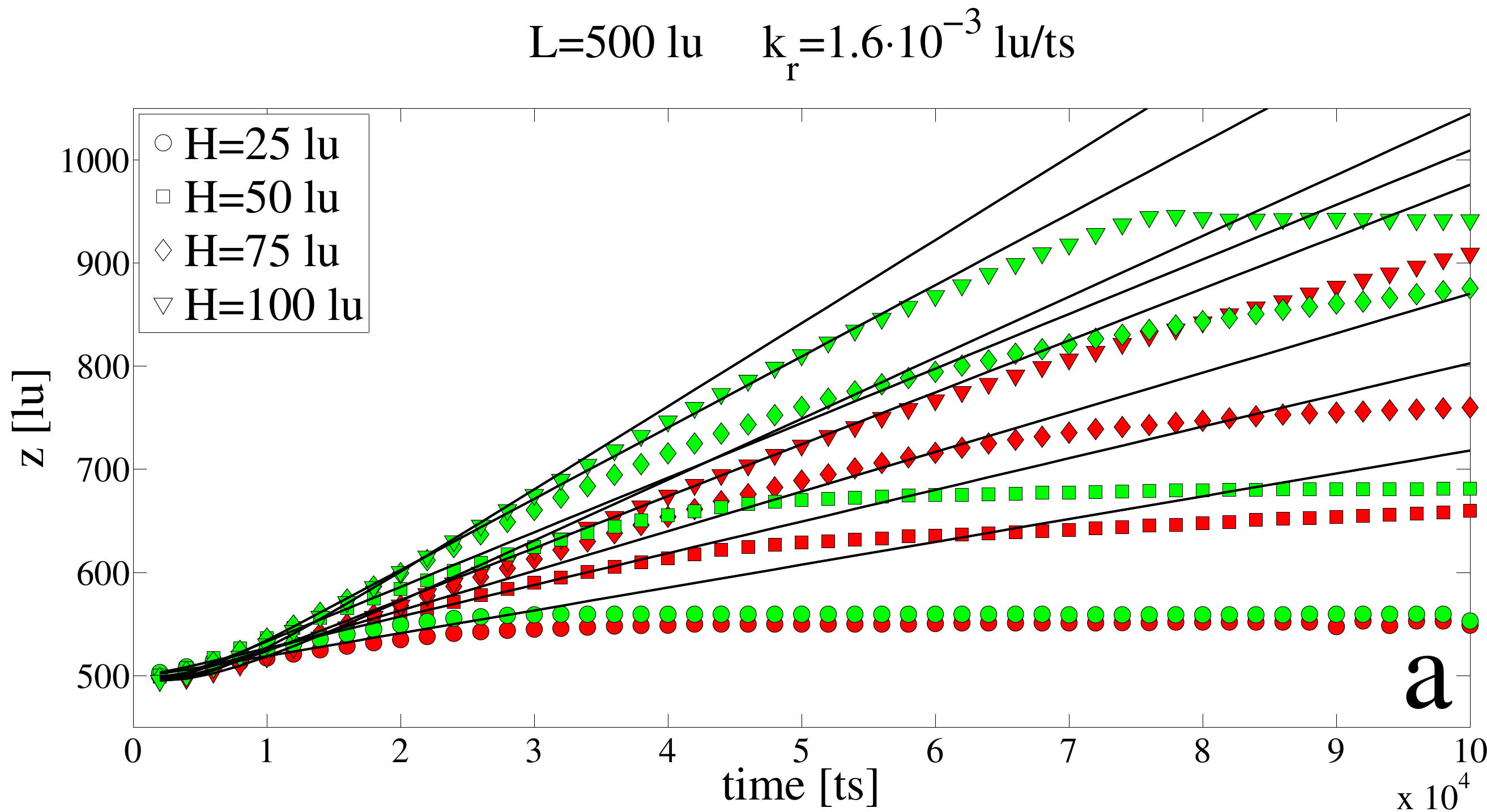}\hspace{0.5cm}
\includegraphics[width=8.5cm]{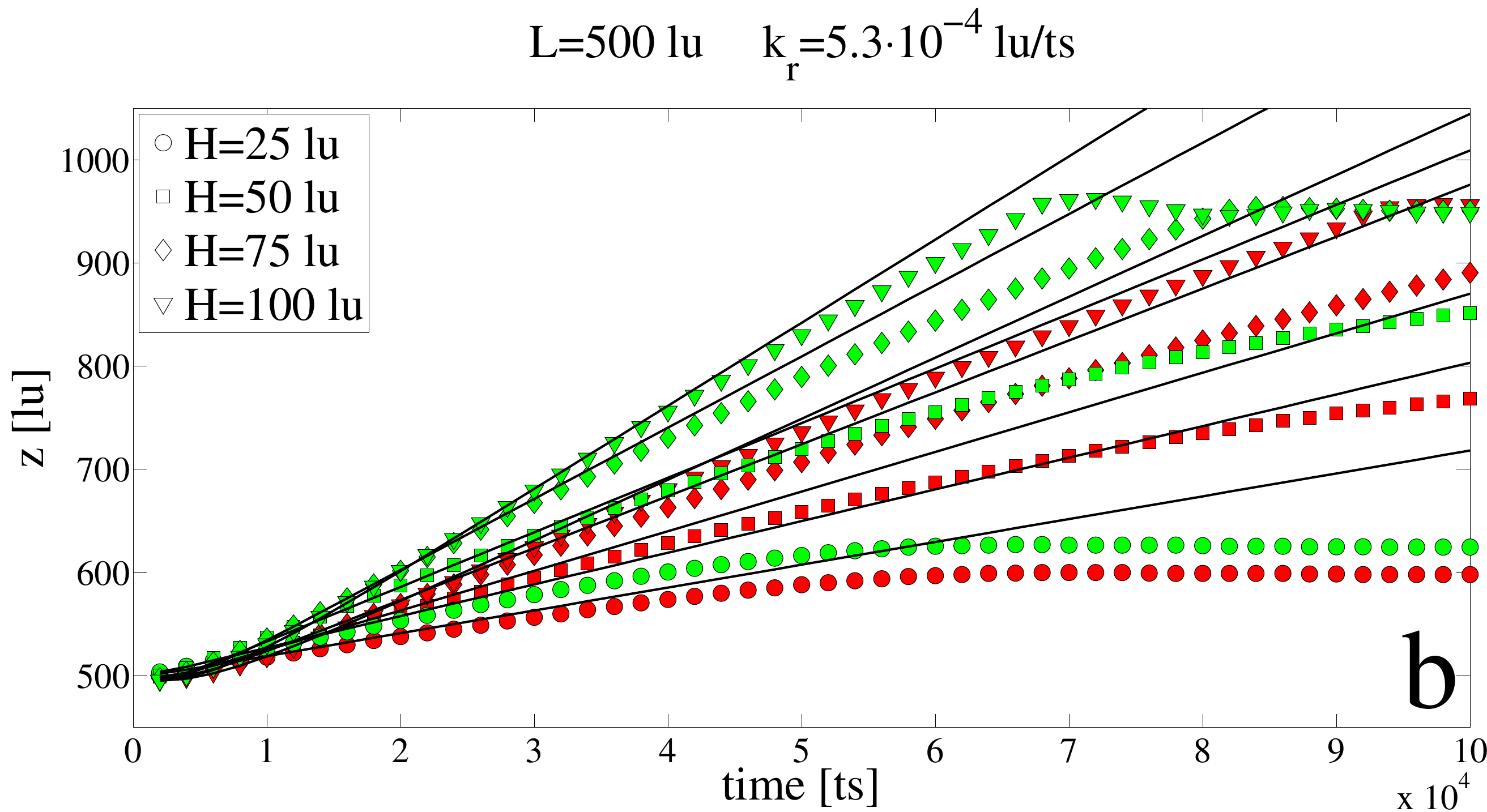}\\
\includegraphics[width=8.5cm]{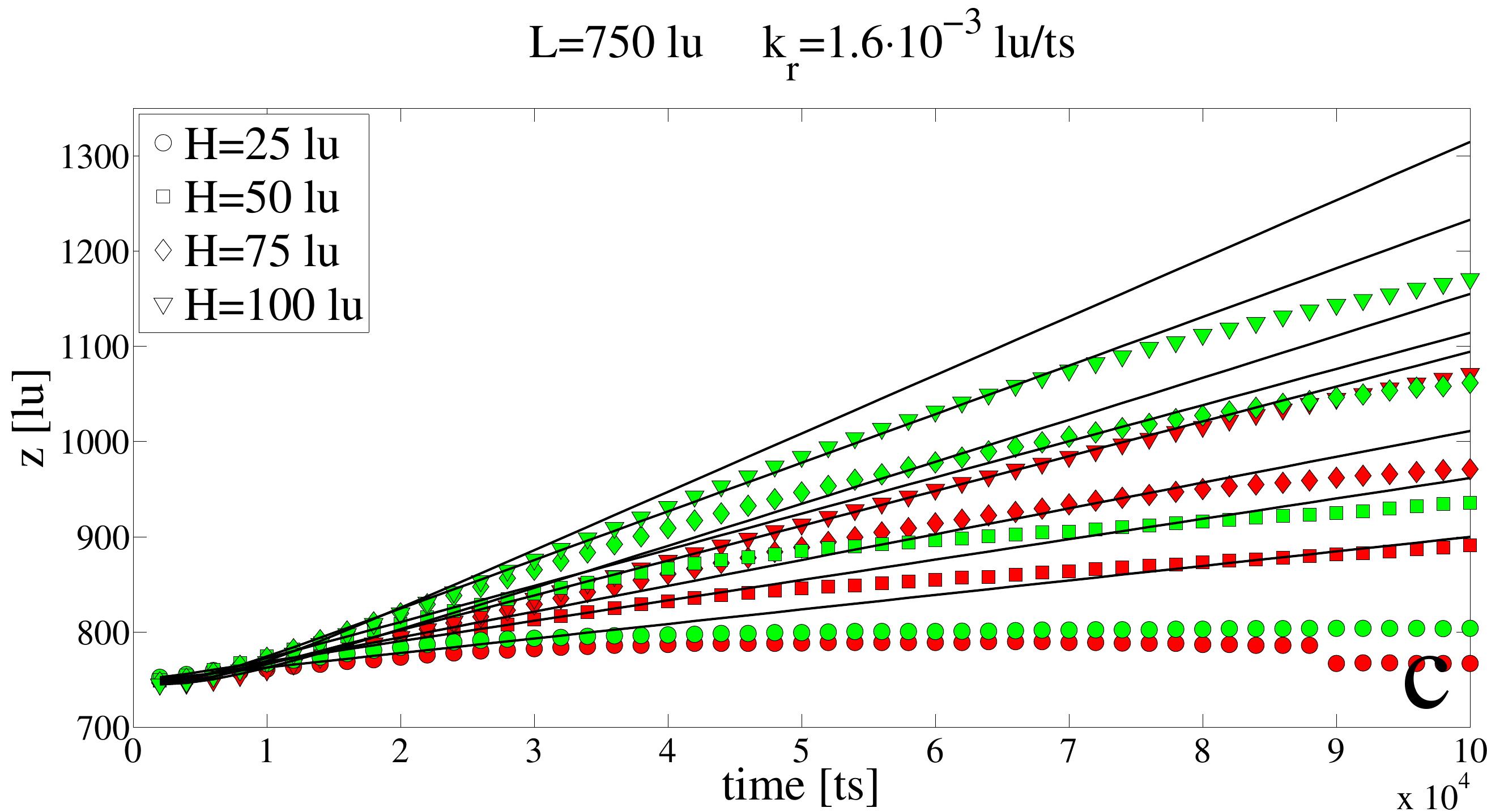}\hspace{0.5cm}
\includegraphics[width=8.5cm]{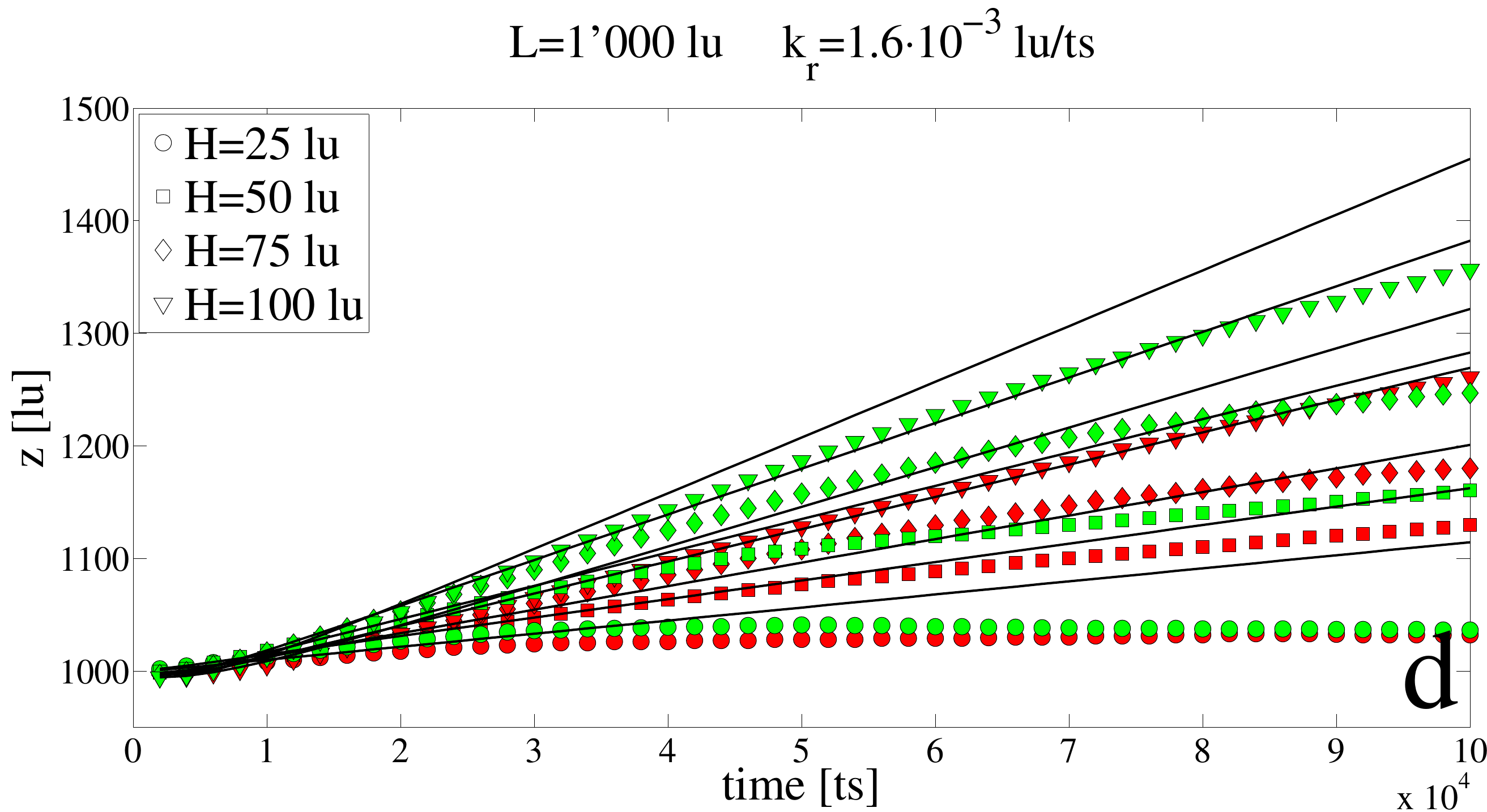}
\caption{\label{fig:front_reaction}
Time dependence of the infiltration depth with reactive boundaries. Data in
red correspond to the surface tension $\gamma=0.12338$ lu$\cdot$mu/ts$^{2}$ ($G_{\mathrm{c}}=0.8$ lu/mu/ts$^{2}$). 
The data represented in green refer to the surface tension $\gamma=0.16403$ lu$\cdot$mu/ts$^{2}$ 
($G_{\mathrm{c}}=0.9$ lu/mu/ts$^{2}$). The solid lines indicate the theoretical predictions in the 
absence of surface growth, Eq.~\ref{eq:washburn}.}
\end{figure*}
\begin{figure*}[t]
\includegraphics[width=8.5cm]{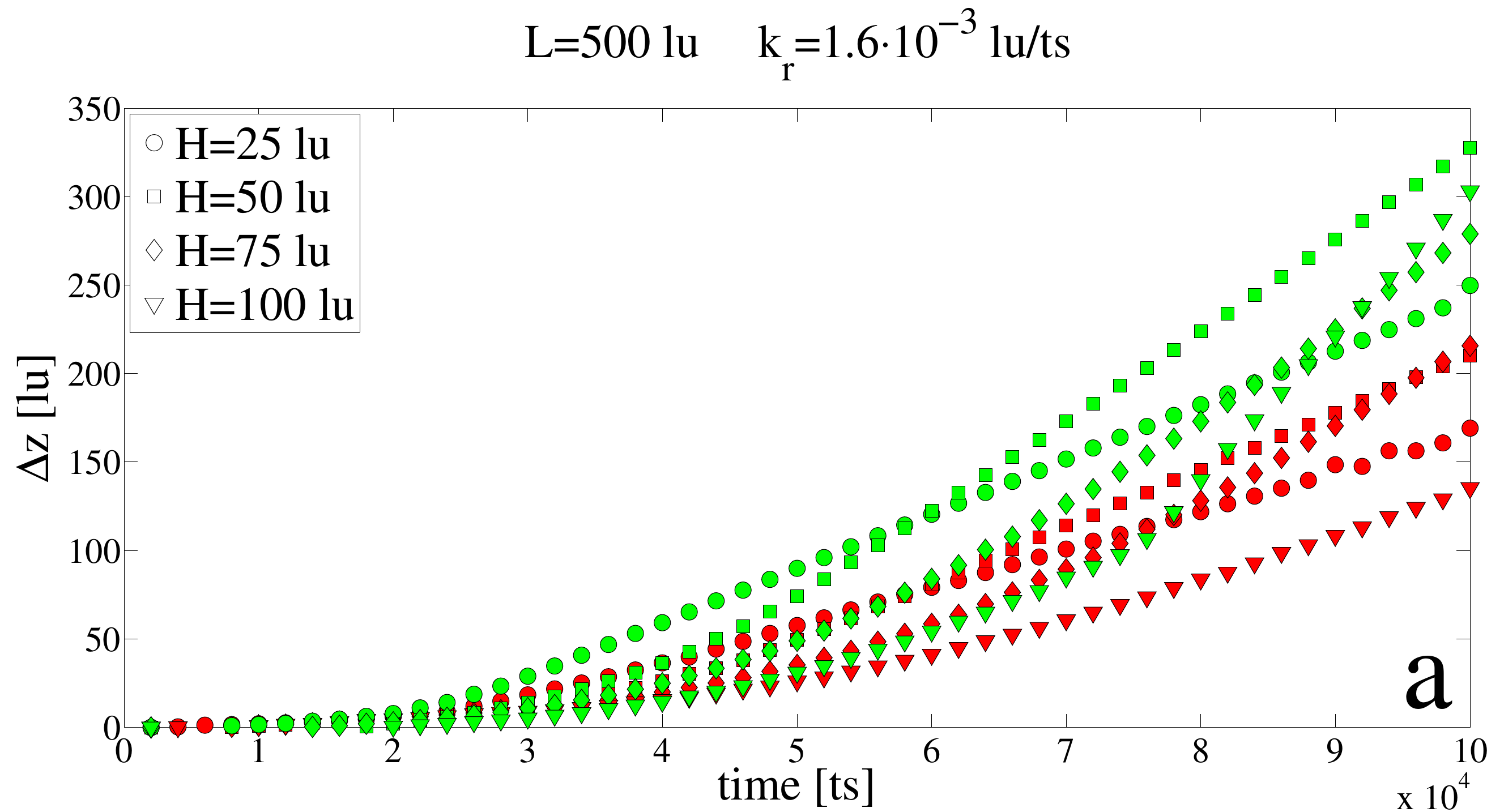}\hspace{0.5cm}
\includegraphics[width=8.5cm]{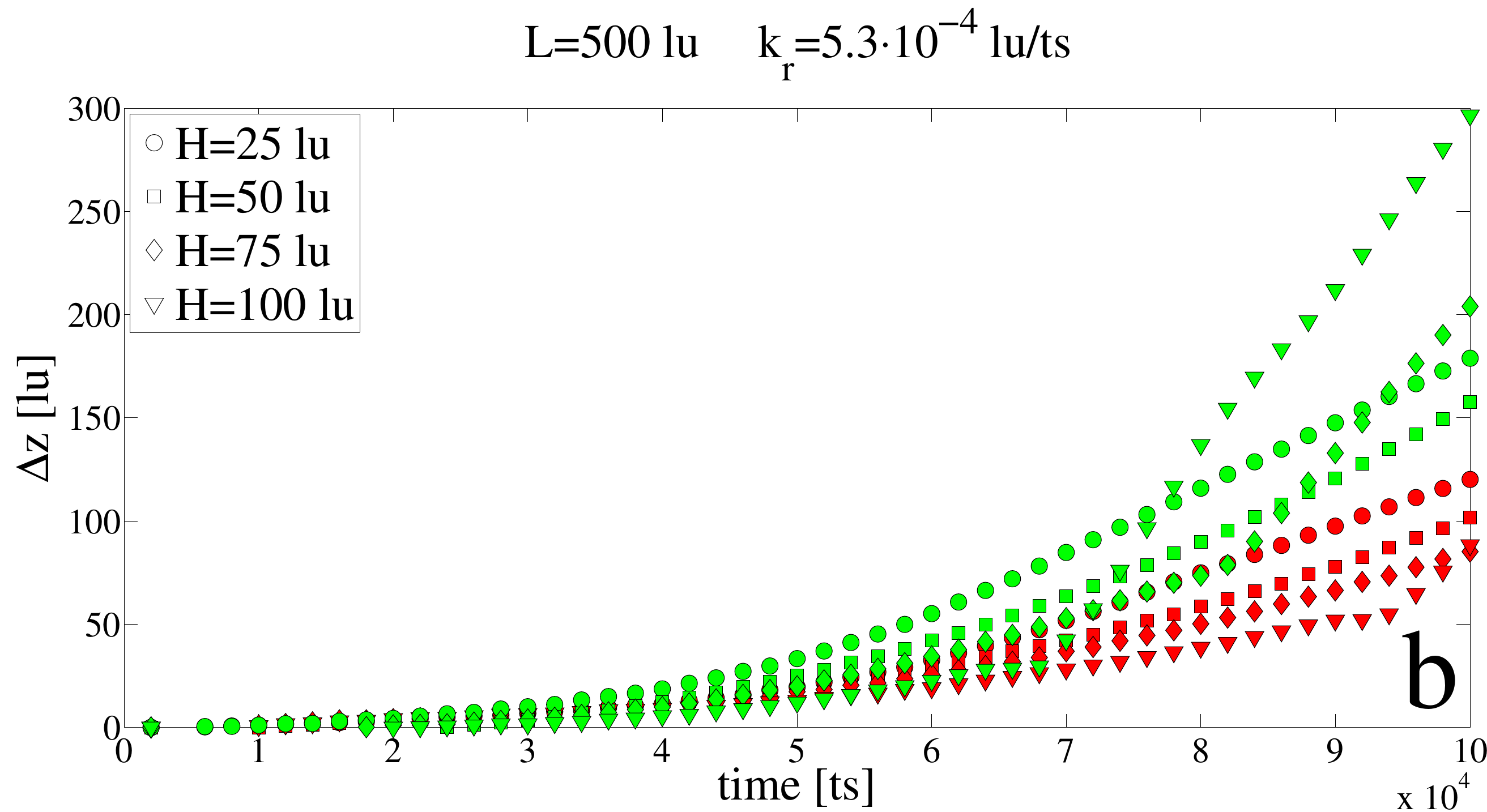}\\
\includegraphics[width=8.5cm]{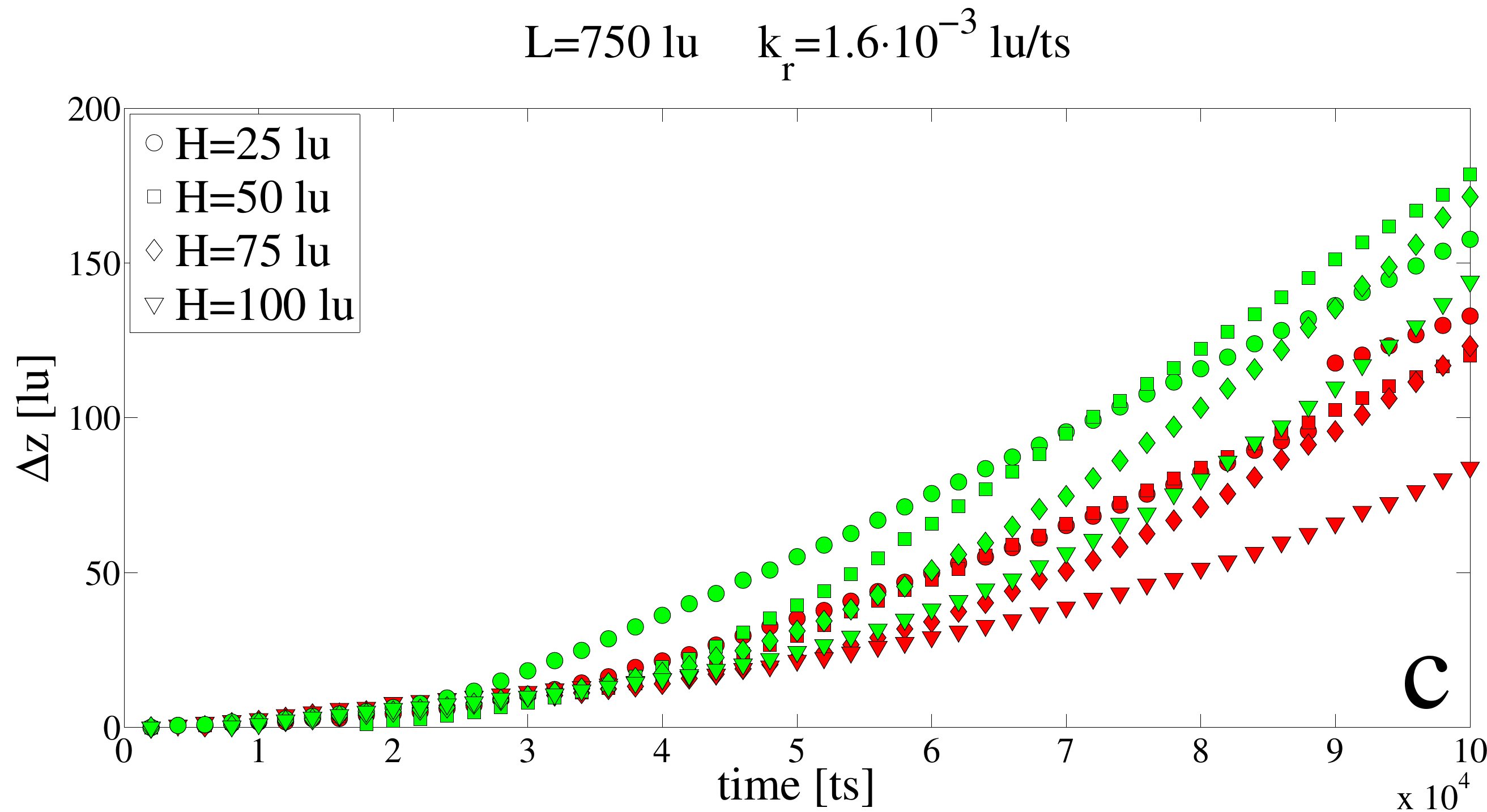}\hspace{0.5cm}
\includegraphics[width=8.5cm]{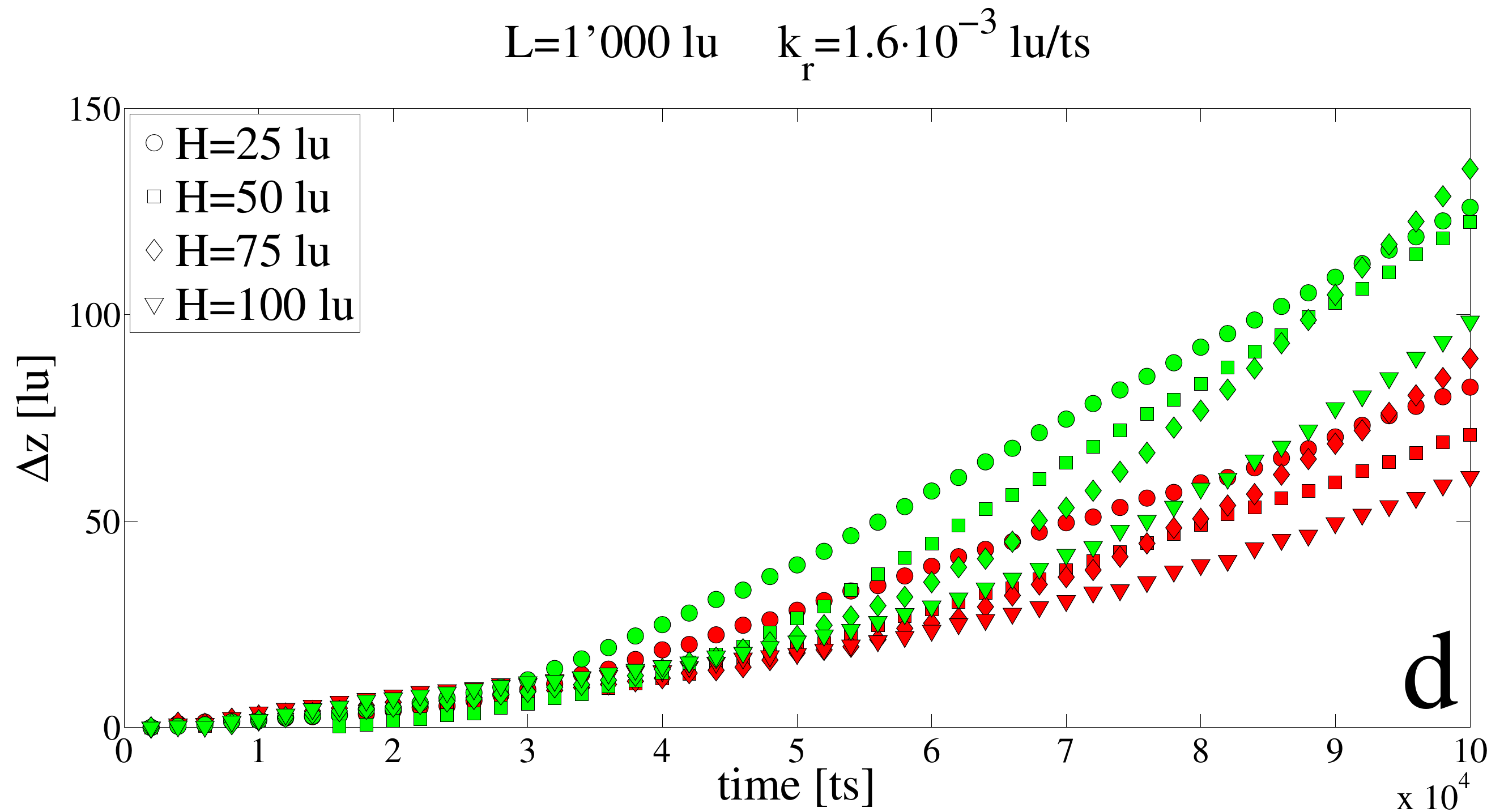}
\caption{\label{fig:dz_reaction}
$\Delta z=z_{\mathrm{t}}-z_{\mathrm{r}}$ in the course of time. The lower script $\mathrm{t}$
refers to the theoretical value without reaction, Eq.~\ref{eq:washburn}; the symbol
$\mathrm{r}$ indicates instead the front displacement for reactive boundaries.
Red for data with surface tension $\gamma=0.12338$ lu$\cdot$mu/ts$^{2}$ ($G_{\mathrm{c}}=0.8$ lu/mu/ts$^{2}$); 
green for data with surface tension $\gamma=0.16403$ lu$\cdot$mu/ts$^{2}$ ($G_{\mathrm{c}}=0.9$ lu/mu/ts$^{2}$).
It should be noted that linearity indicates that pore closure occurred in the case of reactive boundaries. This 
phenomenon is more frequent for narrow interstices. Another clear cause of linearity resides in the length of 
the capillary, that is, when the fluid reaches the other extremity in the absence of surface growth.}
\end{figure*}
\begin{figure*}[t]
\includegraphics[width=8.5cm]{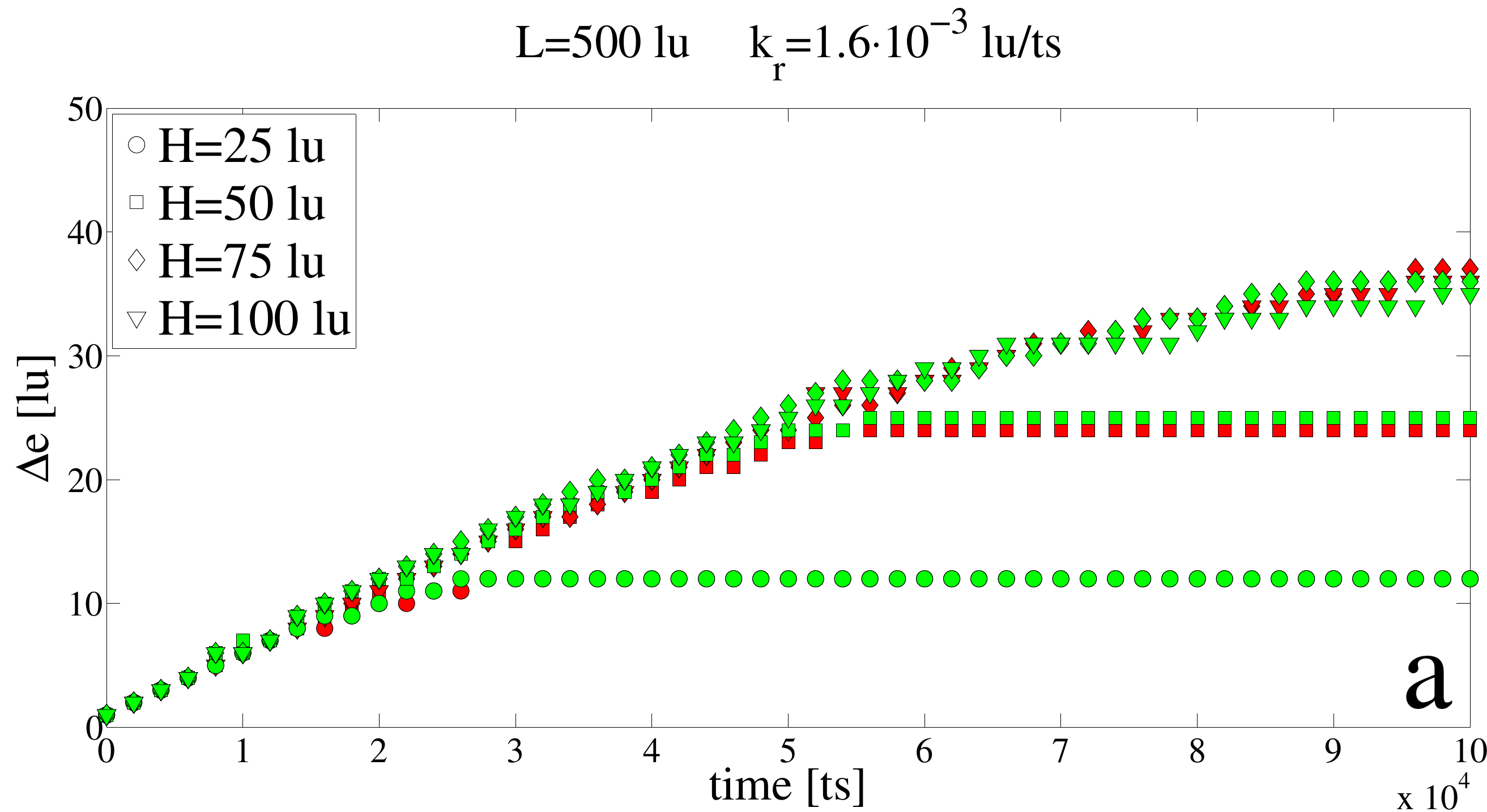}\hspace{0.5cm}
\includegraphics[width=8.5cm]{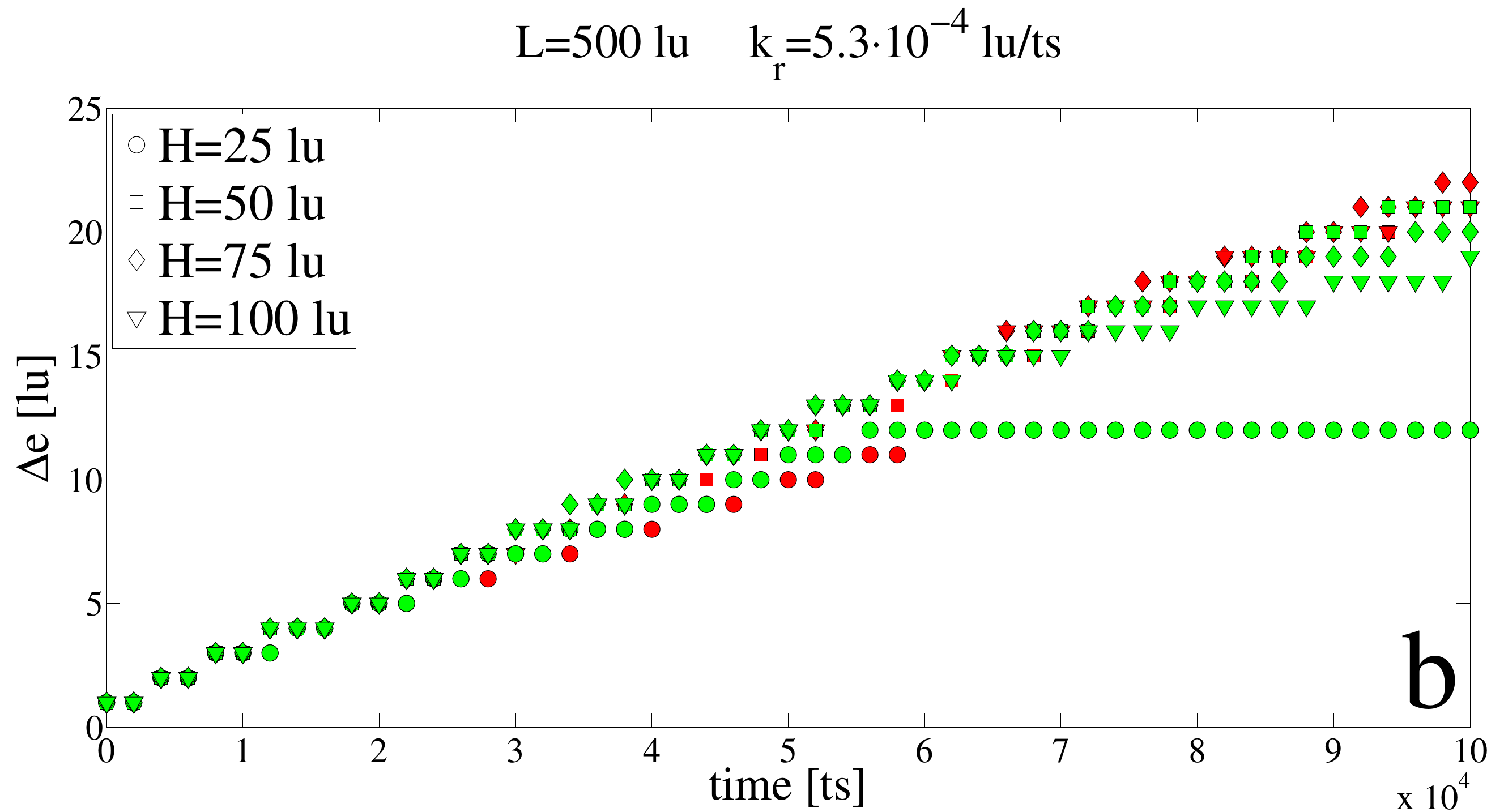}\\
\includegraphics[width=8.5cm]{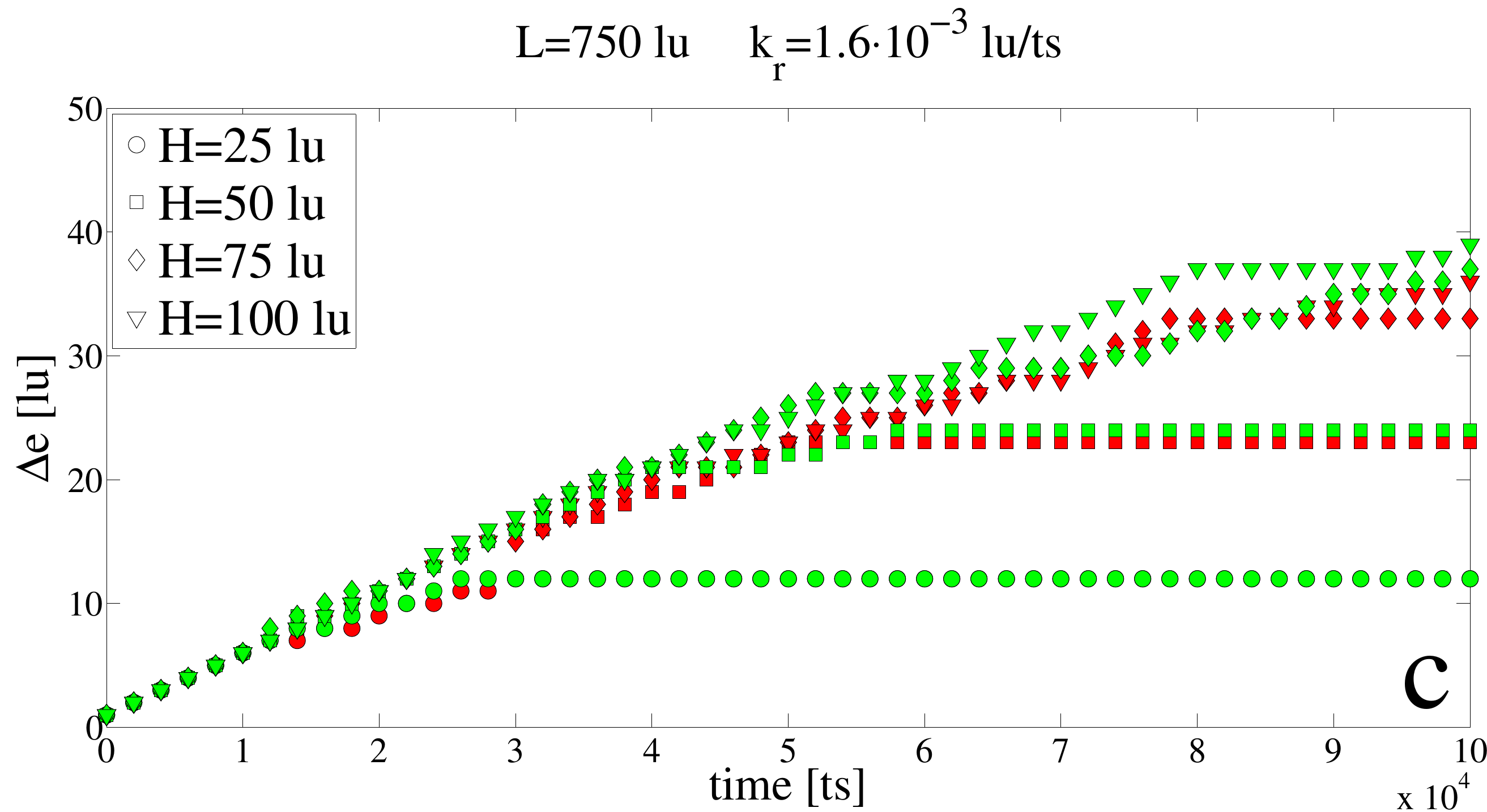}\hspace{0.5cm}
\includegraphics[width=8.5cm]{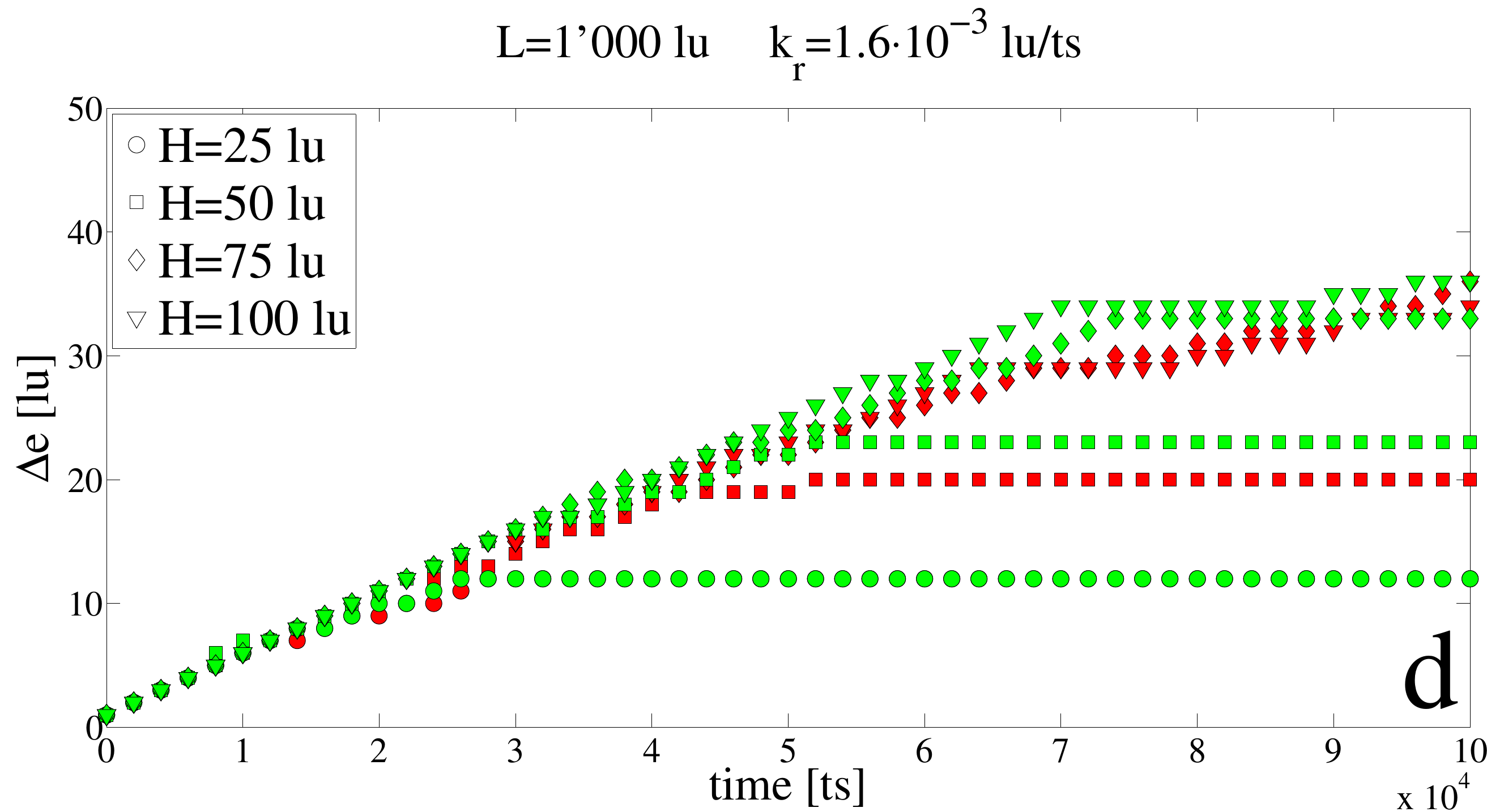}
\caption{\label{fig:solid_reaction}
Time evolution of the maximal width $\Delta e$ of the solid phase inside the capillary. Data in red have surface 
tension $\gamma=0.12338$ lu$\cdot$mu/ts$^{2}$ ($G_{\mathrm{c}}=0.8$ lu/mu/ts$^{2}$); data in green have surface 
tension $\gamma=0.16403$ lu$\cdot$mu/ts$^{2}$ ($G_{\mathrm{c}}=0.9$ lu/mu/ts$^{2}$).}
\end{figure*}
\begin{figure*}[t]
\includegraphics[width=8.5cm]{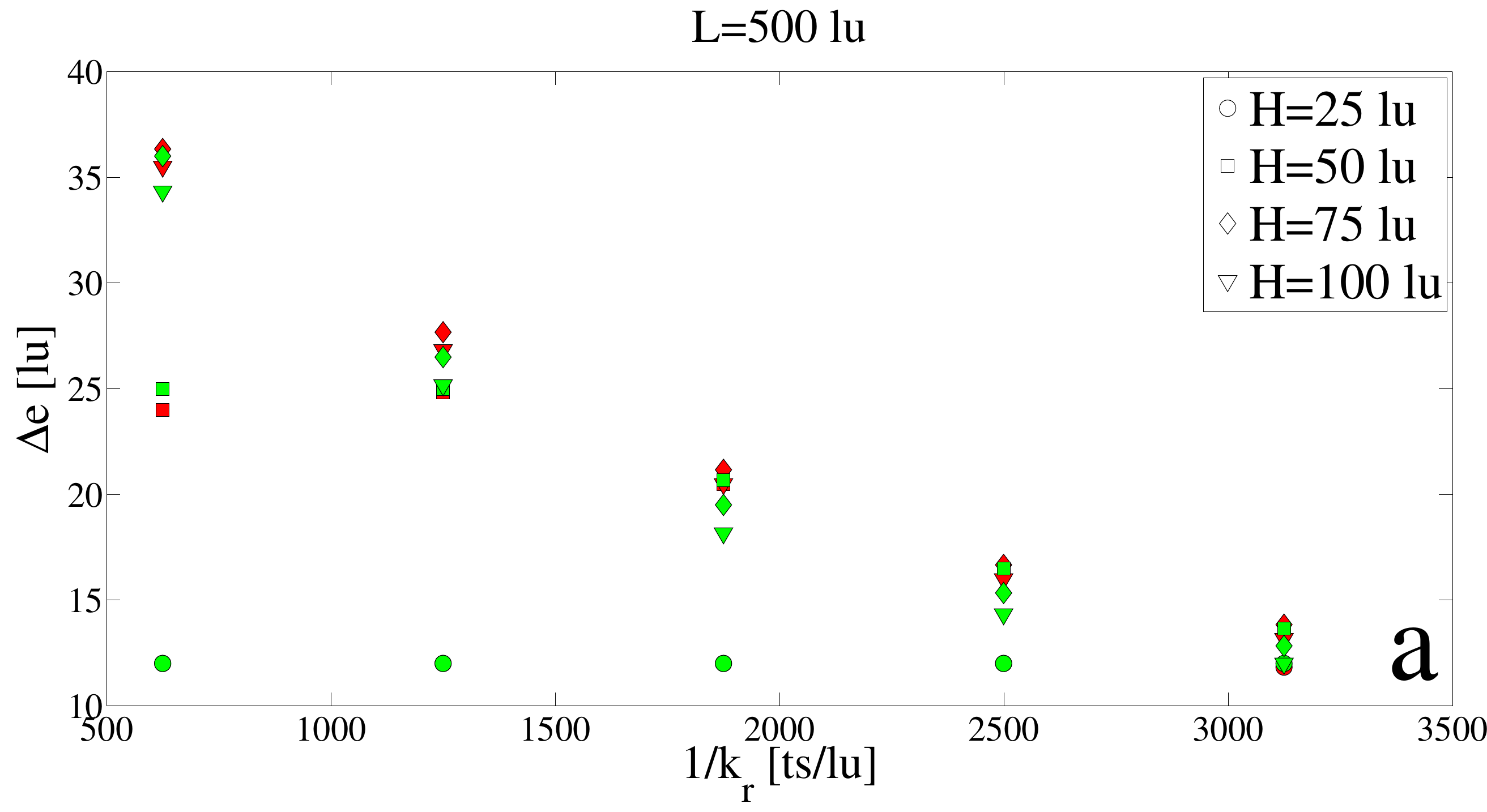}\hspace{0.5cm}
\includegraphics[width=8.5cm]{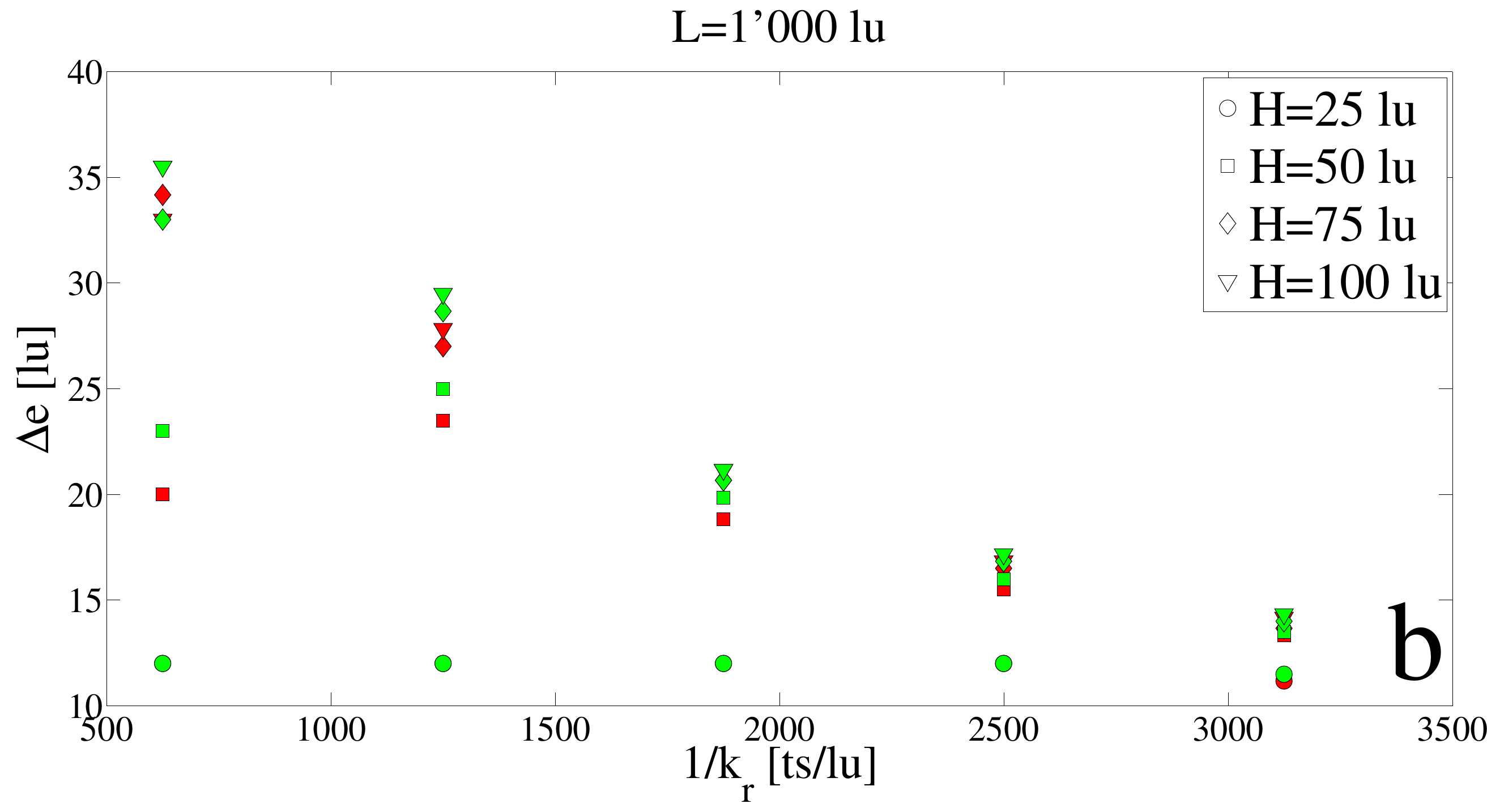}
\caption{\label{fig:solid_kr}
Maximal width $\Delta e$ of the solid phase inside the capillary as a function of the reaction-rate constant $k_{\mathrm{r}}$
averaged over the last five frames. Data in red corresponds to the surface tension $\gamma=0.12338$ lu$\cdot$mu/ts$^{2}$ 
($G_{\mathrm{c}}=0.8$ lu/mu/ts$^{2}$); data in green are associated with the surface tension $\gamma=0.16403$ lu$\cdot$mu/ts$^{2}$ 
($G_{\mathrm{c}}=0.9$ lu/mu/ts$^{2}$).}
\end{figure*}

\subsection*{3.3~~~Capillary flow}\label{sec:washburn}

Capillary flow occurs when a liquid spontaneously spreads inside a channel.
This phenomenon was exhaustively modeled by Washburn and Lukas (Lukas, 1918; Washburn, 1921). 
It may be mentioned that historically capillarity played
also an important role in the development of molecular theories (Israelachvili, 2011).
For two fluids with the same density and dynamic viscosity, by taking 
into account inertial effects, the centerline position of the invading front $z(t)$ 
satisfies the differential equation
(Szekely et al., 1971)
\begin{equation}
\frac{\mathrm{d}^{2}z(t)}{\mathrm{d}t^{2}}+\frac{12\mu}{H^{2}\rho}\frac{\mathrm{d}z(t)}{\mathrm{d}t}
=\frac{2\gamma\cos\theta}{H\rho L}\ .
\label{eq:diff}
\end{equation}
The first term on the left-hand side accounts for the inertial resistance and the second one
for the viscous resistance. The term on the right-hand side is due to capillary forces.
$H$ and $L$ are the height and length of the capillary. $\theta$ is the contact angle
formed by the wetting fluid. Under the assumption of a constant contact angle, a
solution of Eq.~\ref{eq:diff} is obtained by (Chibbaro et al., 2009a)
\begin{equation}
z(t)=\frac{V_{\mathrm{cap}}H\cos\theta}{6L}t_{\mathrm{d}}\big[\exp\big(-t/t_{\mathrm{d}}\big)+t/t_{\mathrm{d}}
-1\big]+z_{0}\ .
\label{eq:washburn}
\end{equation}
The initial position of the interface is denoted by $z_{0}$. $V_{\mathrm{cap}}=\gamma/\mu$ is
the capillary speed and $t_{\mathrm{d}}=\rho H^{2}/12\mu$ is a typical transient time (Chibbaro et al., 2009a).

For times sufficiently large, Eq.~\ref{eq:washburn} leads to $z(t)\propto t$ and the
infiltration rate is given by $K=V_{\mathrm{cap}}H\cos\theta/6L$. Thus, this model reproduces
correctly the linear dependence of the displacement front with time observed for molten Si
and some alloys in porous carbon (Bougiouri et al., 2006; Calderon et al., 2010a; Voytovych et al., 2008). 
Nevertheless, some hypothesis on which the result of Eq.~\ref{eq:washburn} resides are not fulfilled in the
experimental systems. For example, the liquid and vapor phases do not have approximately the 
same density. In the LB model, the density of both fluids determine the surface tension
(see Eqs.~\ref{eq:laplace} and \ref{eq:pressure}). In the sequel, we will focus primarily 
on systems having an equilibrium contact angle around $30^{\circ}$ (see Sec.~3.2). 
This result is consistent with the experimental results for molten
Si and alloys on porous carbon substrates (Bougiouri et al., 2006; Voytovych et al., 2008). Instead, the 
hypothesis assuming the same viscosity for both fluids is more restricting. Indeed, the attempt to increase 
the ratio between the liquid and vapor phases in the LB model would result in a 
breakdown of the linear Washburn law (Chibbaro, 2008). In that respect, it is worthy of mention
the fact that for Si alloys, without reaction, spontaneous infiltration does not even occur (Bougiouri et al., 2006).
It follows that the present model only reproduces the macroscopic desired behavior and the effect 
of the reaction at the contact line is equivalent to the presence, in the capillary, of a second 
fluid as viscous as the wetting component. As a consequence, this means that the role
of the surface reaction between Si and C to form SiC can not be reduced to a simple
transition from a non-wetting to a wetting regime at the contact line of the invading 
front.

In order to give evidence of the behavior predicted by Eq.~\ref{eq:washburn}, we consider
systems similar to that shown in Fig.~\ref{fig:capillary}. As before, the initial densities
satisfy $\rho_{1}/\rho_{2}=2.5\%$ with $\rho_{0}=\rho_{1}+\rho_{2}=2$ mu/lu$^{2}$. The width of 
the capillary varies from $25$ to $100$ lu, corresponding to the height of the simulation domain.
The length of the simulation domain is instead set initially to $1'000$ lu. The length of the
capillary is $L=500$ lu. The left extremity of the capillary coincides with $x=450$ lu.
Every system is let evolve for $100'000$ timesteps. The dynamics is studied by collecting 
$50$ evenly-spaced frames. In the analysis of the data, we first track the invading front along the 
centerline of the capillary, by employing the method described in Sec.~3.1. The contact
angle $\theta$ in Eq.~\ref{eq:washburn} is derived from the capillary pressure using the formula
$\Delta P=2\gamma\cos\theta/H$ (Chibbaro, 2008; Chibbaro et al., 2009a). In this case, the pressures 
$P_{\mathrm{in}}$ and $P_{\mathrm{out}}$ are computed by averaging the pressure $P(\bm{r})$, 
Eq.~\ref{eq:pressure}, over 
regions of the type $[0,2L_{0}+1]\times[0,2L_{0}+1]$. These regions are centered around a point $20$ lu 
apart from the centerline position of the meniscus in both the wetting and non-wetting fluids. It is 
assumed that $L_{0}=10$ lu. This procedure yields the most accurate matching between theory and
simulation results (Chibbaro et al., 2009a), allowing also to avoid difficulties related to the proper definition
and determination of the profile near the contact line (Huang et al., 2007). Figure \ref{fig:front} 
shows the invading front in the course of time for the system of Fig.~\ref{fig:capillary}. Agreement with the 
theoretical prediction is satisfactory, as shown in Fig.~\ref{fig:washburn}. Of course, the surface tension 
remains unchanged since it is not a property depending on the geometry of the system. It turns out that 
the dynamic contact angle is on average larger than the equilibrium contact angle. Figure \ref{fig:thetavsH} 
indicates that the dynamic contact angle increases with $H$. Nevertheless, the process of infiltration is still faster 
for the capillaries of larger width (see Fig.~\ref{fig:washburn}). The present simulations are also repeated
for $L=750$ lu ($N_{x}=1'500$ lu) and $L=1'000$ lu ($N_{x}=2'000$ lu) using similar systems. The results 
are not presented for brevity. The analysis leads to the same conclusions discussed so far.


\subsection*{3.4~~~Capillary infiltration with reactive boundaries}

The SiC formation at the interface is responsible for wetting and spontaneous 
infiltration (Bougiouri et al., 2006). This process is not accounted for directly since our systems are composed by a
wetting and a non-wetting fluid, separated by an interface. This means that in the present 
modeling scheme the reaction at the triple line and related phenomena are not taken into account
(Calderon et al., 2010b; Dezellus et al., 2005; Voytovych et al., 2008). Furthermore, solute transport does not affect fluid flow.
Since the fluid and the solute represent the same substance, it follows
that fluid flow is influenced only by surface growth and not by the
diffusion and precipitation processes. It is thus assumed that the amount of Si involved
in the surface reaction is negligible with respect to the part regarded as a liquid.

The role of reactivity is investigated using again the set-up illustrated in Fig.~\ref{fig:capillary}
in the course of evolutions totaling $100'000$ timesteps.
Regarding fluid flow,  we consider all the systems of Sec.~3.3
with the same initial conditions and the same settings defining the different
wetting behaviors. For solute transport, in the initial
condition, the region occupied by the wetting fluid is filled with solute
at the concentration of $C_{1}=10^{-2}$ mu/lu$^{2}$. In the region occupied 
by the non-wetting fluid, the solute concentration is set to $C_{2}=2\cdot 10^{-3}$ 
mu/lu$^{2}$. For the parameters of $\bm{F}_{\mathrm{s}}$, Eq.~\ref{eq:fs}, we choose 
$G_{\mathrm{s}}=-4.875\cdot 10^{-3}$ mu/lu/ts$^{2}$, $\varphi_{0}=1$ and $C_{0}=4.9\cdot 10^{-3}$ 
mu/lu$^{2}$. The saturated concentration is quite small compared to the bulk value of 
solute: it is chosen to be $C_{\mathrm{s}}=5\cdot 10^{-3}$ mu/lu$^{2}$, determining the relative
saturation $\psi=C_{1}/C_{\mathrm{s}}=2$. As a consequence, for relatively small values 
of $k_{\mathrm{r}}$, the wetting fluid can be regarded as a reservoir for solute (Mortensen et al., 1997). 
Furthermore, these settings guarantee that in the non-wetting fluid, no solute deposits on the 
surface. The initial mass on solid boundaries is $b_{0}=2\cdot 10^{-3}$ mu, and the threshold
value for surface growth is assumed to be $b_{\mathrm{max}}=10^{-2}$ mu. In order to adjust the effect 
of reaction relative 
to diffusion, it is sufficient to change the reaction-rate constant $k_{\mathrm{r}}$, without varying
the relaxation time for solute transport, set to $\tau_{\mathrm{s}}=1$ ts (Kang et al., 2003 and 2004).
Indeed, the systems can be classified according to the relative saturation $\psi=C_{1}/C_{\mathrm{s}}$
and the Damkohler number $Da=k_{\mathrm{r}}N_{y}/D$. The reaction-rate constant is varied
according to the rule $k_{\mathrm{r}}=(8/5)/T$ with $T=1'000,2'000,\dots,5'000$ ts. Indicatively,
this means that the flat surface start growing after $1'000,2'000,\dots,5'000$
timesteps. Figure \ref{fig:growth} shows a typical configuration. It can be seen that the solid 
phase grows especially near the capillary throat. As evidenced by Fig.~\ref{fig:front_reaction}, the
invading front varies linearly with time also for  reactive boundaries. We see that by varying
$\gamma$, $k_{\mathrm{r}}$ and $L$ the relative position between the various curves appears to
be the same. This means that the hierarchy between the associated infiltration rates remains
unchanged. It is of course necessary to take into account that narrow interstices obstruct
sooner. In Fig.~\ref{fig:dz_reaction} we compare more closely the position of the invading front 
to the theoretical prediction in the absence of surface reaction, Eq.~\ref{eq:washburn}. The
retardation in terms of distance is obtained by considering $\Delta z=z_{\mathrm{t}}-z_{\mathrm{r}}$; the
lower scripts $\mathrm{t}$ and $\mathrm{r}$ stand for theoretical and reaction, specifying the two
types of data. First of all, regarding the modeling approach, it is interesting to see that the
curves corresponding to different surface tensions are separated more neatly. For the higher surface
tension, more distance divides the invading front with reaction to that without reaction. Another
general remark is that, for the wider and longer interstices, the retardation in terms of distance
tends to increase faster and almost linearly towards the end of the simulations. According to this
representation of the simulation results, long interstices seem to be the optimal configuration,
but this is not true. For example, for $\gamma=0.12338$ lu$\cdot$mu/ts$^{2}$ and $H=100$ lu,
the invading front spans the distance of $400$ lu if $L=500$ lu as opposed to approximately
$300$ lu if $L=1'000$ lu. Let us now consider the maximal width $\Delta e$ of the solid phase 
in the channel. This quantity can not exceed $N_{y}/2$. This value corresponds to pore closure. 
Figure \ref{fig:solid_reaction} substantiates the discussion in Kang et al.~(2003) related to 
the effect of the fluid velocity on boundary reactivity. It arises that these curves primarily 
depend on the reaction-rate constant $k_{\mathrm{r}}$. The surface tension $\gamma$ and the length 
of the capillary $L$ do not seem to affect significantly the quantity  $\Delta e$. Curves for 
interstices with different width $H$ differ appreciably only when pore obstruction occurs.
As a result, the process of surface growth leading to pore closure appears to be independent of the 
infiltration velocity. Narrow capillaries are thus more subjected to closure and their infiltration
velocities are also smaller. It follows that narrow interstices are particularly detrimental for 
reactive fluid flow. Figure \ref{fig:solid_kr} shows the maximal width of the solid phase resulting
from surface growth as a function of the reaction-rate constant $k_{\mathrm{r}}$ with an average taken
over the last five frames. Also with this representation it can be seen that in passing
from $L=500$ lu to $L=1'000$ lu the data are grouped almost similarly, even though
the velocity decreases significantly. The largest difference is observed for the systems with
$k_{\mathrm{r}}=1.6\cdot 10^{-3}$ lu/ts, $H=50$ lu and $\gamma=0.12338$ lu$\cdot$mu/ts$^{2}$. For
$L=1'000$ lu, the quantity $\Delta e$ only attains $20$ lu. We consider this simulation an 
outlier. From Fig.~\ref{fig:solid_reaction} we see that for $L=750$ lu pore closure occurs with 
$\Delta e=23$ lu as for $L=500$ lu.


\section*{4.~~~Conclusions}

In the present work we have studied through LB simulations in 2D the capillary
infiltration in a single interstice with reactive boundaries.
The complex process of liquid Si infiltration in C preforms was at the center of
our attention. Severe inconsistencies exist between experimental and numerical 
conditions. 
For example, our models only reproduce the expected macroscopic behavior for the infiltration.
This means that the linear dependence on time arising from the reactivity 
(Bougiouri et al., 2006; Israel et al., 2010; Voytovych et al., 2008)
is equivalent to the presence of a second fluid with the same density and viscosity as the reactive 
infiltrant component. In other words, our systems only reproduce the effective hydrodynamic behavior
for the wetting fluid. In addition, our simulations overestimate the typical Reynolds numbers 
(Einset, 1996; Voytovych et al., 2008). 
In principle, the experimental conditions could be met with long channels. But from Fig.~\ref{fig:solid_reaction} we see that the length has little to no effect when the infiltration velocity 
is already small. In this work the focus is on a single capillary, but for interconnected porous systems 
the infiltration process turns out to be slower (Einset, 1996). Another simplifying assumption is that 
the LB simulations are performed in the isothermal regime. It has been demonstrated 
that SiC formation is a highly exothermic reaction (Sangsuwan et al., 1999); moreover, spreading
and infiltration appear to be quite sensitive to the temperature (Bougiouri et al., 2006). As a 
consequence, quantitative agreement is poor and it seems difficult to pinpoint 
the origin of inaccuracy given the mutual dependence of the various complex 
mechanisms at play. Furthermore, the Damkohler number $Da$ is a free parameter in
our simulations. In that respect, for interconnected porous systems, one way to establish
a contact with the experimental results could be to consider the SiC formation at the
droplet contact area when flow stops. The structural properties of the resulting surface
should allow to choose more accurately the range of interest for the parameters 
controlling the reaction. Nevertheless, the proposed analysis still provides us 
with some guidance. In a real porous preform the pores have different sizes and thus
the Damkohler number $Da$ varies locally. This means that the SiC compound does not
grow in the same way in all interstices. So, all our systems are representative for
the problems of flow retardation and pore closure. It is found that the thickening behind the 
invading front effectively hinders the process of capillary infiltration. It turns out 
that wide and short interstices can limit the
effect of pore closure (see Figs.~\ref{fig:front_reaction} and \ref{fig:dz_reaction}).
In other words, porous pathways as straight as possible should ease and dominate
infiltration. Concerning the industrial manufacture of C/SiC composites (Krenkel, 2005), 
it follows that the most
extended surface of the carbon matrix should be put in contact with molten Si,
of course, if the porosity is isotropic.
Then, surface growth does not result in a uniform corrugation of the
initial flat surface but concentrates near the throat (see Fig.~\ref{fig:growth}).
Thus, another way to reduce the slow-down of reactive capillary infiltration could consist
in having round-shaped morphologies for the structure of the porous medium. 
Importantly, narrow interstices are doubly disadvantageous since the process of pore closure 
can be regarded as independent of the infiltration velocity of fluid flow 
(see Fig.~\ref{fig:solid_reaction}). In that respect, it is worth noticing that the benefits 
of bimodal distributions for pore size have been recognized in experimental investigations (Ortona et al., 2012).
Similar conclusions were also reached in the seminal work of Messner and Chiang (1990). In their
theoretical study, the invading front has a parabolic time dependence and the effects of surface reaction
are introduced in an ad-hoc way (cf.~Bougiouri et al.~(2006)). Our next step would be to relate more 
tightly the coupled phenomena of the evolution of pore structure and capillary infiltration to 
the parameters controlling the porosity.
The LB method has the advantage to offer a pore-scale description. This capability can enable
an improvement of the manufacturing conditions of ceramic structures. To this end, possible 
extensions of our work may account for the structural features of single channels, pore
connectivity and the packing structures of particles reproducing the porosity of 
the matrix.


\acknowledgments

The research leading to these results has received funding from the European
Union Seventh Framework Programme (FP7/2007-2013) under grant agreement
n$^{\circ}$ 280464, project "High-frequency ELectro-Magnetic technologies
for advanced processing of ceramic matrix composites and graphite expansion''
(HELM).

\section*{References}

\begin{itemize}
\item[1.]
Benzi R, Succi S, Vergassola M (1992). 
The lattice Boltzmann equation: theory and applications.
\textit{Physics Reports} 222(3):145-197.

\item[2.]
Bhatnagar P, Gross E, Krook A (1954). 
A model for collision processes in gases. I. Small amplitudes processes in charged and neutral one-component systems.
\textit{Physical Review} 94(3):511-525.

\item[3.]
Bougiouri V, Voytovych R, Rojo-Calderon N, Narciso J, Eustathopoulos N (2006). 
The role of the chemical reaction in the infiltration of porous carbon by NiSi alloys.
\textit{Scripta Materialia} 54(11):1875-1878.

\item[4.]
Calderon NR, Voytovych R, Narciso J, Eustathopoulos N (2010a). 
Pressureless infiltration versus wetting in AlSi/graphite systems.
\textit{Journal of Materials Science} 45(16):4345-4350.

\item[5.]
Calderon NR, Voytovych R, Narciso J, Eustathopoulos N (2010b).
Wetting dynamics versus interfacial reactivity of AlSi alloys on carbon.
\textit{Journal of Materials Science} 45(8):2150-2156.

\item[6.]
Chen S, Doolen GD (1998). 
Lattice Boltzmann method for fluid flows.
\textit{Annual Review of Fluid Mechanics} 30:329-364.

\item[7.]
Chibbaro S (2008). 
Capillary filling with pseudo-potential binary Lattice-Boltzmann model.
\textit{The European Physical Journal E} 27(1):99-106.

\item[8.]
Chibbaro S, Biferale L, Diotallevi F, Succi S (2009a).
Capillary filling for multicomponent fluid using the pseudo-potential Lattice Boltzmann method.
\textit{The European Physical Journal Special Topics} 171(1):223-228.

\item[9.]
Chibbaro S, Costa E, Dimitrov DI, Diotallevi F, Milchev A, Palmieri D,
Pontrelli G, Succi S (2009b). Capillary filling in microchannels with wall corrugations: A comparative
study of the Concus-Finn criterion by continuum, kinetic, and atomistic approaches. 
\textit{Langmuir} 25(21):12653-12660.

\item[10.]
Dezellus O, Eustathopoulos N (2010).
Fundamental issues of reactive wetting by liquid metals.
\textit{Journal of Materials Science} 45(16):4256-4264.

\item[11.]
Dezellus O, Hodaj F, Eustathopoulos N (2003).
Progress in modelling of chemical-reaction limited wetting.
\textit{Journal of the European Ceramic Society} 23(15):2797-2803.

\item[12.]
Dezellus O, Jacques S, Hodaj F, Eustathopoulos N (2005). 
Wetting and infiltration of carbon by liquid silicon.
\textit{Journal of Materials Science} 40(9-10):2307-2311.

\item[13.]
Diotallevi F, Biferale L, Chibbaro S, Lamura A, 
Pontrelli G, Sbragaglia M, Succi S, Toschi T (2009a).
Capillary filling using lattice Boltzmann equations: The case of multi-phase flows.
\textit{The European Physical Journal Special Topics} 166(1):111-116.

\item[14.]
Diotallevi F, Biferale L, Chibbaro S, Pontrelli G, Toschi F, Succi S (2009b). 
Lattice Boltzmann simulations of capillary filling: Finite vapour density effects.
\textit{The European Physical Journal Special Topics} 171(1):237-243.

\item[15.]
Einset EO (1996). 
Capillary infiltration rates into porous media with application to silcomp processing.
\textit{Journal of the American Ceramic Society} 79(2):333-338.

\item[16.]
Einset EO (1998). 
Analysis of reactive melt infiltration in the processing of ceramics and ceramic composites.
\textit{Chemical Engineering Science} 53(5):1027-1039.

\item[17.]
Eustathopoulos N, Nicholas MG, Drevet B (1999).
\textit{Wettability at High Temperatures}, Pergamon.

\item[18.]
de Gennes PG, Brochard-Wyart F, Qu\'er\'e D (2004).
\textit{Capillarity and Wetting Phenomena: Drops, Bubbles, Pearls, Waves},
Springer.

\item[19.]
He X, Luo LS (1997). 
A priori derivation of the lattice Boltzmann equation.
\textit{Physical Review E} 55(6):6333-6336.

\item[20.]
Hillig WB, Mehan RL, Morelock CR, DeCarlo VJ, Laskow W (1975).
Silicon/silicon carbide composites.
\textit{American Ceramic Society Bulletin} 54(12):1054-1056.

\item[21.]
Huang H, Thorne DT, Shaap MG, Sukop MC (2007). 
Proposed approximation for contact angles in Shan-and-Chen-type multicomponent multiphase lattice Boltzmann models.
\textit{Physical Review E} 76(6):66701-66706.

\item[22.]
Israel R, Voytovych R, Protsenko P, Drevet B, Camel D, Eustathopoulos N (2010). 
Capillary interaction between molten silicon and porous graphite.
\textit{Journal of Materials Science} 45(8):2210-2217.

\item[23.]
Israelachvili JN (2011). 
\textit{Intermolecular and Surface Forces}, Academic Press.

\item[24.]
Kang Q, Lichtner PC, Zhang D (2007). 
An improved lattice Boltzmann model for multicomponent reactive transport in porous media at the pore scale.
\textit{Water Resources Research} 43:5551-5562.

\item[25.]
Kang Q, Zhang D, Chen S (2003).
Simulation of dissolution and precipitation in porous media.
\textit{Journal of Geophysical Research} 108(B10):2504-2513.

\item[26.]
Kang Q, Zhang D, Chen S (2002a).
Displacement of a two-dimensional immiscible droplet in a channel.
\textit{Physics of Fluids} 14(9):3203-3214.

\item[27.]
Kang Q, Zhang D, Chen S, He X (2002b). 
Lattice Boltzmann simulation of chemical dissolution in porous media.
\textit{Physical Review E} 65(3):36318-36325.

\item[28.]
Kang Q, Zhang D, Lichtner PC, Tsimpanogiannis IN (2004).
Lattice Boltzmann model for crystal growth from supersaturated solution.
\textit{Geophysical Research Letters} 31:21107-21111.

\item[29.]
Krenkel W (2005). 
Carbon fibre reinforced silicon carbide composites (C/SiC, C/C-SiC).
\textit{Handbook of Ceramic Composites}, Kluwer Academic Publisher, 117-48.

\item[30.]
Landau LD, Lifshitz EM (2008). \textit{Fluid Mechanics}, Elsevier.

\item[31.]
Liu GW, Muolo ML, Valenza F, Passerone A (2010).
Survey on wetting of SiC by molten metals.
\textit{Ceramics International} 36(4):1177-1188.

\item[32.]
Lu G, DePaolo DJ, Kang Q, Zhang D (2009). 
Lattice Boltzmann simulation of snow crystal growth in clouds.
\textit{Journal of Geophysical Research} 114:11087-11100.

\item[33.]
Lucas R (1918). 
Rate of capillary ascension of liquids.
\textit{Kooloid-Z} 23:15-22.

\item[34.]
Martys NW, Chen HD (1996). 
Simulation of multicomponent fluids in complex three-dimensional geometries by lattice 
Boltzmann method. \textit{Physical Review E} 53(1):743-750.

\item[35.]
Miller W, Succi S (2002). 
A lattice Boltzmann model for anisotropic crystal growth from melt.
\textit{Journal of Statistical Physics} 107(1-2):173-186.

\item[36.]
Messner RP, Chiang YM (1990). 
Liquid-phase reaction-bonding of silicon carbide using alloyed silicon-molybdenum melts.
\textit{Journal of the American Ceramic Society} 73(5):1193-1200.

\item[37.]
Mohamad AA, El-Ganaoui M, Bennacer R (2009).
Lattice Boltzmann simulation of natural convection in an open ended cavity.
\textit{International Journal of Thermal Sciences} 48(10):1870-1875.

\item[38.]
Mohamad AA, Kuzmin A (2010).
A critical evaluation of force term in lattice Boltzmann method, natural convection problem.
\textit{International Journal of Heat and Mass Transfer} 53(5-6):990-996.

\item[39.]
Mortensen A, Drevet B, Eustathopoulos N (1997).
Kinetics of diffusion-limited spreading of sessile drops in reactive wetting.
\textit{Scripta Materialia} 36(6):645-651.

\item[40.]
Ortona A, Fino P, D'Angelo C, Biamino S, D'Amico G, Gaia D, Gianella S (2012). 
Si-SiC-ZrB$_{2}$ ceramics by silicon reactive infiltration.
\textit{Ceramics International} 38(4):3243-3250.

\item[41.]
Pooley CM, Kusumaatmaja H, Yeomans JM (2009).
Modelling capillary filling dynamics using lattice Boltzmann simulations.
\textit{The European Physical Journal Special Topics} 171(1):63-71.

\item[42.]
Sangsuwan P, Tewari SN, Gatica JE, Singh M, Dickerson R (1999).
Reactive infiltration of silicon melt through microporous amorphous carbon preforms.
\textit{Metallurgical and Materials Transaction B} 30B(5):933-944.

\item[43.]
Schmieschek S, Harting J (2011).
Contact angle determination in multicomponent lattice Boltzmann simulations.
\textit{Communications in Computational Physics} 9(5):1165-1178.

\item[44.]
Sergi D, Scocchi G, Ortona A (2012). 
Molecular dynamics simulations of the contact angle between water droplets and graphite surfaces.
\textit{Fluid Phase Equilibria} 332:173-177.

\item[45.]
Shan X, Chen H (1993). 
Lattice Boltzmann model for simulating flows with multiple phases and components.
\textit{Physical Review E} 47(3):1815-1819.

\item[46.]
Shan X, Chen H (1994).
Simulation of nonideal gases and liquid-gas phase transitions by the lattice Boltzmann equation.
\textit{Physical Review E} 49(4):2941-2948.

\item[47.]
Shan X, Doolen GD (1995). 
Multicomponent lattice-Boltzmann model with interparticle interaction.
\textit{Journal of Statistical Physics} 81(1-2):379-393.

\item[48.]
Succi S (2009). \textit{The Lattice Boltzmann Equation for Fluid Dynamics and Beyond}, 
Oxford University Press.

\item[49.]
Sukop MC, Thorne Jr DT (2010). \textit{Lattice Boltzmann Modeling:
An Introduction for Geoscientists and Engineers}, Springer.

\item[50.]
Szekely J, Neumann AW, Chuang YK (1971). 
The rate of capillary penetration and the applicability of the Washburn equation.
\textit{Journal of Colloid and Interface Science} 35(2):273-278.

\item[51.]
Voytovych R, Bougiouri V, Calderon NR, Narciso J, Eustathopoulos N (2008). 
Reactive infiltration of porous graphite by NiSi alloys.
\textit{Acta Materialia} 56(10):2237-2246.

\item[52.]
Washburn EW (1921). 
The dynamics of capillary rise. \textit{Physical Review} 27(3):273-283.

\item[53.]
Wolf-Gladrow DA (2005). \textit{Lattice-Gas Cellular Automata
and Lattice Boltzmann Models - An Introduction}, Springer.

\item[54.]
Wolf-Gladrow D (1995). 
A lattice Boltzmann equation for diffusion.
\textit{Journal of Statistical Physics} 79(5-6):1023-1032.

\item[55.]
Young T (1805). 
An essay on the cohesion of fluids.
\textit{Philosophical Transactions of the Royal Society of London} 95:65-87.
\end{itemize}
\end{document}